\documentclass[11pt,a4paper]{article}
\usepackage{longtable}
\usepackage{amsmath}
\usepackage{amssymb}
\usepackage{graphicx}
\usepackage{bm}
\usepackage{footmisc}
\usepackage{afterpage}
\usepackage{float}
\usepackage{lscape}
\usepackage{longtable}
\usepackage{float}
\usepackage{slashed}
\usepackage{morefloats}
\usepackage{afterpage}

\usepackage{definitions}
\usepackage{xspace}
\usepackage{xcolor}
\usepackage{heppennames2}

\newcommand*{\no}{\noindent}
\newcommand*{\bea}{\begin{eqnarray}}
\newcommand*{\eea}{\end{eqnarray}}
\newcommand*{\be}{\begin{equation}}
\newcommand*{\ee}{\end{equation}}

\newcommand*{\pref}[1]{(\ref{#1})}

\newcommand*{\prefr}[2]{(\ref{#1}-\ref{#2})} 
\newcommand*{\nn}{\nonumber}

\newcommand{\bma}{\begin{pmatrix}}
\newcommand{\ema}{\end{pmatrix}}

\newcommand*{\op}{{{\cal O}}}
\newcommand*{\la}{\left\langle}
\newcommand*{\ra}{\right\rangle}

\newcommand{\delphes} {{\textsc{Delphes}}\xspace}

\newcommand{\ttZ}{\mbox{\ensuremath{\symbolFace{t}\overline{\symbolFace{t}}\cPZ}}\xspace}
\newcommand{\ttW}{\mbox{\ensuremath{\symbolFace{t}\overline{\symbolFace{t}}\cPW}}\xspace}

\newcommand{\wjets}{\mbox{\ensuremath{\symbolFace{W}}+jets}\xspace}

\def\lep{\mbox{\ensuremath{\symbolFace{l}}}\xspace}
\def\Zoff{\mbox{\ensuremath{\symbolFace{l}^+\symbolFace{l}^-}}\xspace}
\def\Wplep{\mbox{\ensuremath{\symbolFace{l}^+\symbolFace{\nu}_{\symbolFace{l}}}}\xspace}
\def\Wmlep{\mbox{\ensuremath{\symbolFace{l}^-\tilde{\symbolFace{\nu}}_{\symbolFace{l}}}}\xspace}

\def\ttbar{\mbox{\ensuremath{\symbolFace{t}\overline{\symbolFace{t}}}}\xspace}

\def\WZ{\mbox{\ensuremath{\symbolFace{W}\symbolFace{Z}}}\xspace}
\def\tWZ{\mbox{\ensuremath{\symbolFace{t}\symbolFace{W}\symbolFace{Z}}}\xspace}

\def\tW{\mbox{\ensuremath{\symbolFace{t}\symbolFace{W}}}\xspace}
\def\tZ{\mbox{\ensuremath{\symbolFace{t}\symbolFace{Z}}}\xspace}
\def\Zg{\mbox{\ensuremath{\symbolFace{Z}\gamma}}\xspace}

\newcommand{\jet}{\ensuremath{\symbolFace{j}}}

\newcommand{\pp}{\ensuremath{\symbolFace{p}\symbolFace{p}}}

\usepackage[square,comma,sort&compress,numbers]{natbib}

\title{Constraining the Higgs valence contribution\\in the proton\footnote{Report number UWTHPH-19-38.}}
\small
\author{Simon Fernbach$^1$, Lukas Lechner$^2$, Axel Maas$^1$,\\ Simon Pl\"atzer$^3$, Robert Sch\"ofbeck$^{2}$\\[0.5cm]
$^1$Institute of Physics, NAWI Graz, University of Graz,\\
Universit\"atsplatz 5, A-8010 Graz, Austria\\
$^2$Institute of High Energy Physics, Austrian Academy of Sciences,\\Nikolsdorfergasse 18, 1050 Vienna, Austria \\
$^3$University of Vienna,\\Boltzmanngasse 5, A-1090 Vienna, Austria and\\
Erwin Schr\"odinger International Institute for Mathematical Physics (ESI)\\
Boltzmanngasse 9, A-1090 Vienna, Austria}
\normalsize

\begin{document}

\maketitle

\begin{abstract}

Non-perturbative gauge-invariance under the strong and the weak interactions dictates that the proton contains a non-vanishing valence contribution from the Higgs particle.
By introducing an additional parton distribution function~(PDF), we investigate the experimental consequences of this prediction. 
The Herwig 7 event generator and a parameterized CMS detector simulation are used to obtain predictions for a scenario amounting to the LHC Run II data set.
We use those to assess the impact of the Higgs PDF on the $\pp\to \ttbar$ process in the single lepton final state. Comparing to nominal simulation we derive expected limits as a function of the shape of the valence Higgs PDF. We also investigate the process $\pp\to \ttZ$ at the parton level to add further constraints.

\end{abstract}

\section{Introduction}

In gauge theories, physical states must be gauge-invariant.
This simple requirement is implemented perturbatively by the BRST symmetry \cite{Bohm:2001yx,'tHooft:1979bj}. 
Beyond perturbation theory, however, effects such as the Gribov-Singer ambiguity obstruct this construction~\cite{Gribov:1977wm,Singer:1978dk,Fujikawa:1982ss}.
Nevertheless, valid observables correspond to non-perturbatively manifest gauge-invariant operators. 
In non-Abelian gauge theories, these are necessarily composite~\cite{Banks:1979fi,Frohlich:1980gj,Frohlich:1981yi}.
In theories with a gauge structure different to the standard model~(SM), this requirement can change the phenomenology qualitatively~\cite{Maas:2015gma,Maas:2017xzh,Sondenheimer:2019idq}.
This prediction is further corroborated by lattice simulations~\cite{Maas:2016ngo,Maas:2018xxu}, and a review can be found in Ref.~\cite{Maas:2017wzi}.
In the SM, however, effects on phenomenology are expected to be small~\cite{Frohlich:1980gj,Frohlich:1981yi}, again confirmed by lattice simulations~\cite{Maas:2012tj,Maas:2013aia,Egger:2017tkd,Maas:2018ska}.
Small deviations could nevertheless be detectable~\cite{Maas:2012tj,Maas:2018ska,Egger:2017tkd,Maas:2017wzi}, and these are the focus of this work. 

One consequence of these considerations is that, in addition to the perturbative Higgs sea contribution~\cite{Bauer:2017isx,Bauer:2018xag,Bauer:2018arx}, the operator describing the SM proton must necessarily also have a valence Higgs component~\cite{Egger:2017tkd}. 
If the center-of-mass energy exceeds the Higgs boson mass, proton-proton collider phenomenology can be affected~\cite{Egger:2017tkd,Maas:2017wzi}. 
Provided the valence contribution is large enough, a discovery of the effect might be possible at the LHC. 
At the FCC-hh, any effect of the valence Higgs contribution will be greatly amplified\footnote{We note that the valence Higgs implies that only full weak multiplets are present in the proton, which can potentially strongly affect also the sea contribution at FCC-hh \cite{Bauer:2017isx,Bauer:2018xag,Bauer:2018arx}.}.
We discuss the details in Section~\ref{s:theory}

To explore such a possible contribution, we use an ansatz for the valence Higgs PDF and study its impact on the $\pp\to\ttbar$ and $\pp\to \ttZ$ processes  at the parton level.
As detailed in Section~\ref{s:herwig}, this is done by simulating the process with Herwig 7 \cite{Bellm:2015jjp} at the level of the full cross section.

For the process $\pp\to\ttbar$, we also investigate a scenario amounting to the LHC Run-II data set. 
To this end, Section~\ref{s:cms} describes the processing of events with the \delphes~\cite{deFavereau:2013fsa} software package that parametrizes the reconstruction performance of, in our case, the CMS detector~\cite{Chatrchyan:2008aa}. 
We obtain predictions for total cross sections and various differential distributions of discriminating observables at the detector level.
Including sensible estimates of the experimental uncertainties, we derive expected upper limits at 95\%~CL of typically $\mathcal{O}(10^{-4})$ on the total Higgs valence contribution (Section~\ref{s:results}). While the shapes of the Higgs valence PDFs considered in this work only give an indication of potential effects, our results encourage global fits of proton PDF sets including a more general form of the valence Higgs contribution.

A small set of preliminary results has already been made available in Ref.~\cite{Maas:2019dwd}.

\section{The structure of the proton}\label{s:theory}

The requirement of manifest gauge invariance in a non-Abelian gauge theory necessitates composite operators~\cite{Frohlich:1980gj,Frohlich:1981yi}. For the SM, this leads only to minor differences compared to perturbative calculations~\cite{Bohm:2001yx}. See for a detailed review \cite{Maas:2017wzi}, including an overview over support from lattice calculations.

This can be understood as following \cite{Maas:2017wzi}. Consider the observed 125-GeV scalar, viz.\ what is usually called the Higgs. This needs to be described by a gauge-invariant operator. The simplest scalar operator of this type in the standard model is the composite operator
\be
\op_0(x)=(\phi^\dagger\phi)(x)\label{scalar}
\ee
\no where $\phi(x)$ is the full Higgs field. In the usual way it is possible to isolate from the full  Higgs field $\phi$ the fluctuation field $H$ and the vacuum expectation value $v$ \cite{Maas:2017wzi,Bohm:2001yx}.

The scalar particle described by $\op_0$ in \pref{scalar} is a singlet in the fundamental representation of the weak interaction. At the same time an operator like $\op_0$ has the same structure as an operator describing a hadron in QCD. It thus describes a bound state, and would, in principle, require a non-perturbative treatment.

However, the particularities of the Brout-Englert-Higgs effect allows for the treatment of \pref{scalar} in an analytical manner, the Fr\"ohlich-Morchio-Strocchi (FMS) mechanism \cite{Frohlich:1980gj,Frohlich:1981yi}. In a fixed 't Hooft-gauge expand $\phi$ in \pref{scalar} around its vacuum expectation value, and take the connected part only. This yields
\bea
&&\la\op_0^\dagger(x)\op_0(y)\ra_c\label{lhs}\\
&=&v^2\la H(x)H(y)\ra_c\label{term1}\\
&&+v\la H^\dagger(x)H^\dagger(y)H(y)+x\leftrightarrow y\ra_c+\la H^\dagger(x)H(x)H^\dagger(y)H(y)\ra_c\label{term2}.
\eea
\no In perturbation theory, each of the terms is individually BRST-invariant \cite{Bohm:2001yx}. Non-perturbatively, this is spoiled by the Gribov-Singer ambiguity \cite{Maas:2017wzi}, and only the sum is gauge-invariant and physical.

Standard perturbation theory is equivalent to keeping the term \pref{term1}, and expanding it into the perturbative series, while \pref{term2} is discarded.
Viewing that as an expansion in both, $v$ and the usual perturbative series, implies that, to leading order in $v$, the propagator of the bound state coincides with the one of the elementary Higgs in perturbation theory to all orders in the other couplings.
Because the poles coincide, the bound state therefore has the same mass and properties as the elementary Higgs.
Because the contributions \pref{term2} are small and the remainder of the SM works in a very similar way \cite{Frohlich:1980gj,Frohlich:1981yi,Maas:2017wzi},
the approximation is excellent and explains why standard perturbative calculations in the SM are doing so well at LHC. 

However, this does not imply that the remaining terms \pref{term2} must be immeasurably small. 
First investigations on and off the lattice indeed hint that the contributions can become relevant in certain kinematic situations~\cite{Egger:2017tkd,Maas:2018ska,Maas:2019dwd}, though usually energy scales of order $v$ or the Higgs mass are required. 
This suggests to look at the proton as initial state particle at the LHC, as we do in the following.

For the proton, the, entirely group-theoretical, arguments go as follow, see \cite{Egger:2017tkd,Maas:2017wzi} for details:
\begin{itemize}
 \item[1)] Because of the strong interactions, it needs a multiple of three quarks to get a proton which is gauge-invariant under the strong interactions.
 \item[2)] The proton is a parity eigenstate, and can therefore not be constructed from right-handed quarks alone. An odd number of left-handed quarks is necessary.
 \item[3)] The required left-handed quarks carry weak charges, and the proton must hence also be gauge-invariant with respect to to the weak interactions.
 \item[4)] Because the number of weak charges is odd an odd number of fundamental weak charges is needed. In addition, no electric or color charge is allowed to not upset the other quantum numbers.
 \item[5)] This odd number of particles should not alter the spin of the proton, and can thus only be a weak fundamentally charged boson.
\end{itemize}
The Higgs is the only particle in the standard model to fit this description. As an added bonus, the custodial charge carried by the Higgs yields two states, which can be identified with the proton and the neutron. Thus, the Higgs also adds a quantum number to the proton, just like the three valence quarks do. This replaces the ordinary flavor in perturbative descriptions. Thus, consistency requires that the Higgs is treated on the same footing as the valence quarks, rather than to be generated entirely by splitting and evolution. It is also the only particle, besides the three valence quarks, for which this applies. Other particles in the proton, like the photon or the Z boson, are entirely generated from splitting and evolution, as they do not contribute to the quantum numbers of the proton.

Thus, symbolically\footnote{The actual expressions can be found in \cite{Maas:2017wzi,Egger:2017tkd}.}, a proton operator reads
\be
p(x)=(qqq\phi)(x)\label{proton}.
\ee
\no The only external quantum numbers carried by this operator are baryon number and custodial charge, of which its gauged subgroup yields the necessary electric charge. The explicit breaking of the custodial symmetry in the SM creates the differences between neutrons and protons, including their mass difference, up to the electromagnetic corrections.

At leading order in the FMS mechanism $p\approx vqqq$, and thus ordinary QCD arises.
Of course, the operator $qqq$ still needs to be treated non-perturbatively within QCD. 
However, just as in \prefr{lhs}{term2}, additional terms arise beyond leading order. 
In particular, in any scattering process (see \cite{Maas:2017wzi})
\bea
&&\la p(x)p(y)X(z_1,...)\ra=v^2\la (qqq)(y)(qqq)(x)X(z_1,...)\ra\nn\\
&&+v\la (qqq)(x)H(x)(qqq)(y)X(z_1,...)+x\leftrightarrow y\ra\nn\\
&&+\la (qqq)(x)H(x)(qqq)(y)H(y)X(z_1,...)\ra\nn
\eea
\no where $X$ contains all remaining operators. Beyond the leading contribution,  additional terms appear with an explicit Higgs contribution in the initial state.
Because the initial state is now involved, an explicit perturbative calculation, which may be possible for a lepton collider, is probably too complicated within the foreseeable future. 
Thus, the additional effects will be parameterized here, following \cite{Egger:2017tkd}, by an additional valence PDF for the Higgs.

It should be remarked that in many BSM scenario, the same considerations can lead to qualitatively different effects, rather than the small quantitative ones discussed here. Especially, this can lead to qualitatively different spectra and cross sections. See \cite{Maas:2017wzi} for an overview. This strongly motivates to search for the corresponding SM effects to understand the relevance for model building of the underlying field-theoretical considerations.

\section{PDF input}\label{s:herwig}

\subsection{The PDF ansatz}

This leaves to choose the Higgs PDF. As the aim is here a first exploration, we will neglect DGLAP evolution and $Q^2$ dependence, and work with a simple $x$-dependent PDF. The rationale behind this decision is that the two relevant scales are the Higgs mass and the average parton energy in top production at the LHC, which are sufficiently close to each other that the logarithmic running in the evolution will not yield a substantial impact.

In a next step, the valence PDF obtained here could be used, e.\ g., as seed PDF in the evolution equations of \cite{Bauer:2017isx,Bauer:2018xag,Bauer:2018arx}, in which the Higgs seed PDF at the LHC is currently chosen as zero. We defer this to future work. This would be necessary to extend the present investigation to the FCC-hh.

As usual for the valence PDFs, we need to formulate some way of modelling them to constrain it in the following by data. Even though the Higgs is now considered a valence particle, it is expected that some of the condensate features remain. Thus, we expect that the Higgs PDF should be concentrated, to leading order, around $x\approx 0$, similar to the presumed color glass condensate of gluons. This is also in line with the considerations of \cite{Egger:2017tkd}, where the situation at lepton colliders was investigated.

We consider five different forms for the general Higgs PDF $P_H$
\bea
P_H^1(x)&=&(1-x)\exp(-c_t x^2)\label{pdf1}\\
P_H^2(x)&=&\frac{1-x}{x}\exp(-c_t x^2)\label{pdf2}\\
P_H^3(x)&=&\frac{1}{x}\exp(-c_t x^2)\label{pdf3}\\
P_H^4(x)&=&(1-x)\exp\left(-c_t\left(x-\frac{1}{4}\right)^2\right)\label{pdf4}\\
P_H^5(x)&=&\frac{1-x}{x}\exp\left(-c_t\left(x-\frac{1}{4}\right)^2\right)\label{pdf5}.
\eea
\no The parameter $c_t$ is a tuning parameter, which we will tune to reach agreement with the data. Choosing in \prefr{pdf4}{pdf5} a center away from zero at $x=1/4$ yields a scenario where the Higgs is less condensate-like, but more valence-like. Note that for a Higgs to come on-shell requires at least $x\gtrsim 0.02$ at the LHC with a proton energy of 7 TeV. A summary of the optimal tuning parameters $c_t$ obtained in section \ref{ss:selection} is provided as an illustration in Table~\ref{tab:PDFs}.

{\renewcommand{\arraystretch}{1.3}
\begin{table}
\caption{Optimal tuning parameter values for the Higgs PDFs in \prefr{pdf1}{pdf5} based on the selection procedure in section \ref{ss:selection}. The optimal choice for PDF 2 based on the results of section \ref{s:cms} changes $c_t$ to $100$.}
\label{tab:PDFs}
\begin{center}
\begin{tabular}{c|c|c|c|c|c}
PDF & 1 & 2 & 3 & 4 & 5\\\hline
$c_t$ & 4 & 40 & 47 & 37 & 200 \\
\end{tabular}
\end{center}
\end{table} 
}

The presence of the Higgs PDF influences the sum-rules. This is best seen when considering the momentum one \cite{Egger:2017tkd}. 
Including the Higgs leads to a modification of the sum-rule from
\be
\int dx x\left(\sum_{f=g,\bar{q},q} P_f\right)=1\nn,
\ee
\no to
\bea
1&=&\int dx x\left(\sum_{f=g,\bar{q},q} (1-c_f) P_f\right)+c\int dx x P_H^i\nn\\
&=&\int dx x\left(\sum_{f=g,\bar{q},q} (1-c_f)P_f\right)+cf_i(c_t)\nn,
\eea
\no with an explicitly calculable function $f$. 
The parameter $c$ can be viewed as the absolute normalisation of the Higgs PDF $P_H^i$, and describes the 'fraction' of the proton made up by the Higgs. 
The $c_f$ depend on the parton flavor, necessitating in principle a global PDF refit. 
Moreover, the valence Higgs can affect the shape of quark and gluon PDFs as well, or modify their $Q^2$-dependence. 
At the lowest order considered here, however, the gluon will contribute about 90\% of the standard contributions, and thus only $c_g$ plays a substantial role. 
As the Higgs is assumed to form a condensate-like structure while the gluon dominates the QCD-partons at low $x$, a suitable assumption for now will be $c_f=c_g=c$. 
With the usual PDFs fixed, we will thus not be able to perfectly maintain the sum rule, but its violation for reasonable parameter ranges is small.

In the following, we consider primarily the process $pp\to t\overline{t}$, because the top quark has a Yukawa coupling $y_t\approx 1$. The inclusive cross section for this process can be written as
\be
\sigma_{pp\to\ttbar}(c)=(1-c)^2\sigma_{pp\to\ttbar}+(1-c)c\sigma_{gH\to\ttbar}+c^2\sigma_{HH\to\ttbar}\label{totalxs},
\ee
\no where the $c$-independent $\sigma_{pp\to\ttbar}$ contains all ordinary considered contributions in the initial state, i.\ e.\ quarks and gluons.
Initial state Higgs bosons are excluded in this term. 
The experimental uncertainty on the total cross-section will be used in Section \ref{ss:selection} to provide a first constrain of $c$ and the tuning parameters $c_t$ in \prefr{pdf1}{pdf5}.

Furthermore, the kinematics of processes with initial state Higgs and QCD partons differ, and therefore we proceed to differential cross sections, determined as
\bea
&&\frac{d^n\sigma_{pp\to\ttbar}(c)}{\sigma_{pp\to\ttbar}(c)da^1...da^n}=\nn\\
&&\frac{(1-c)^2}{\sigma_{pp\to\ttbar}(c)}\frac{d^n\sigma_{pp\to\ttbar}}{da^1...da^n}+\frac{(1-c)c}{\sigma_{pp\to\ttbar}(c)}\frac{d^n\sigma_{gH\to\ttbar}}{da^1...da^n}\nn\\
&&+\frac{c^2}{\sigma_{pp\to\ttbar}(c)}\frac{d^n\sigma_{HH\to\ttbar}}{da^1...da^n}\label{xs}
\eea
\no with $\sigma_{pp\to\ttbar}(c)$ given by \pref{totalxs} above. Herein the $a_i$ are arbitrary physical observables, like transverse momentum of particles or rapidity.

In addition, we will investigate the process $pp\to \ttZ$ at the level of the full cross section \pref{totalxs} to constrain the tuning parameter $c_t$ in \prefr{pdf1}{pdf5} further.

\subsection{Hard process}

To simulate the process, we use the Herwig 7 event generator
\cite{Bahr:2008pv,Bellm:2015jjp,Bellm:2017bvx} using the Matchbox
module \cite{Platzer:2011bc} and the angular ordered shower
\cite{Gieseke:2003rz}, with matrix elements provided by a combination
of MadGraph5\_aMCatNLO \cite{Alwall:2014hca} and ColorFull
\cite{Sjodahl:2014opa}.

Since a massive Higgs inside the proton cannot be strictly collinear
to the incoming proton's momentum, we allow the Higgs to be space-like
off-shell. To be precise, for a Higgs momentum $p=(p^0,\vec
q_\perp,p^3)$ inside a proton of momentum $P$ we consider $p^2=-\mu^2
< 0$, with a longitudinal momentum fraction
$x=(p^0+p^3)/(P^0 + P^3)$, the Higgs' transverse momentum relates to
its virtuality as \be \vec q_\perp^2=(1-x)\mu^2-(xm_p)^2\ge 0\ . \nn
\ee \no The momentum fraction is then limited in between \be 0\le
x\le\frac{\mu^2}{2m_p^2}\left(\sqrt{1+\frac{4m_p^2}{\mu^2}}-1\right)\nn
\ee \no and we distribute the kinematic variables according to \be
\frac{{\rm d}\mu^2}{\mu^2+m_H^2} \frac{{\rm d}\phi}{2\pi}{\rm d}x \ ,
\nn \ee with a uniform azimuthal orientation $\phi$ of the Higgs
transverse momentum, and an upper bound by the hadronic center of mass
energy $S$ is implied on the transverse momentum scale $\mu^2$. Notice
that in the limit $m_H^2,m_p^2\ll S$ we in fact recover a collinear
Higgs momentum across the entire range of longitudinal momentum
fraction $x$, as the transverse momenta $\mu^2$ are then peaked at
zero. The proton's momentum after the Higgs has been extracted is
simply given by momentum conservation, \be P' = P-p \nn \ee and can be shown to satisfy $(P')^2=m_p^2$ from the kinematics chosen
above.

For the baseline of a standard proton as well as for the QCD parton induced contributions we use
the MMHT2014nlo68cl PDFs \cite{Buckley:2014ana} for quarks and gluons.
The processes involving QCD partons are simulated at NLO with initial states $\bar{q}q$ and
$gg$. As initial states with Higgs content we add gH and
HH. However, at this time NLO calculations with initial-state Higgs
are not supported by Herwig, and we thus evaluate these processes only
at LO.  The results are either evaluated directly at the partonic
level, or showered for input into \delphes using hepmc files.  The
latter is again done at NLO for the QCD parton processes and at LO
for processes with initial state Higgs. The parton showering, as well
as the decays of the top quarks, are then performed with default
settings, and in the case of the Higgs-induced sub-processes no
multi-parton interactions take place.

\begin{figure}
  \begin{center}
    \includegraphics[width=0.5\textwidth]{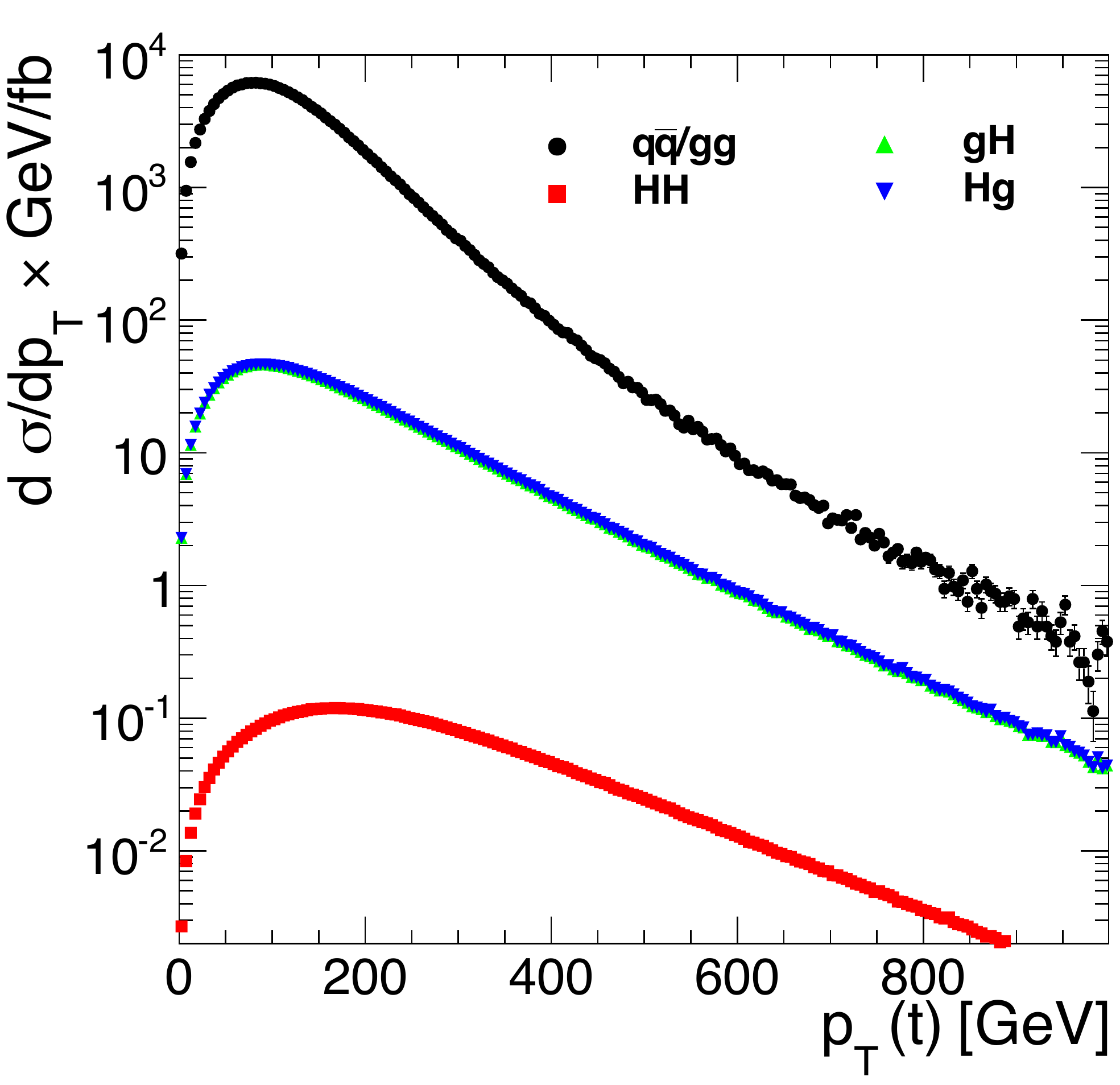}\includegraphics[width=0.5\textwidth]{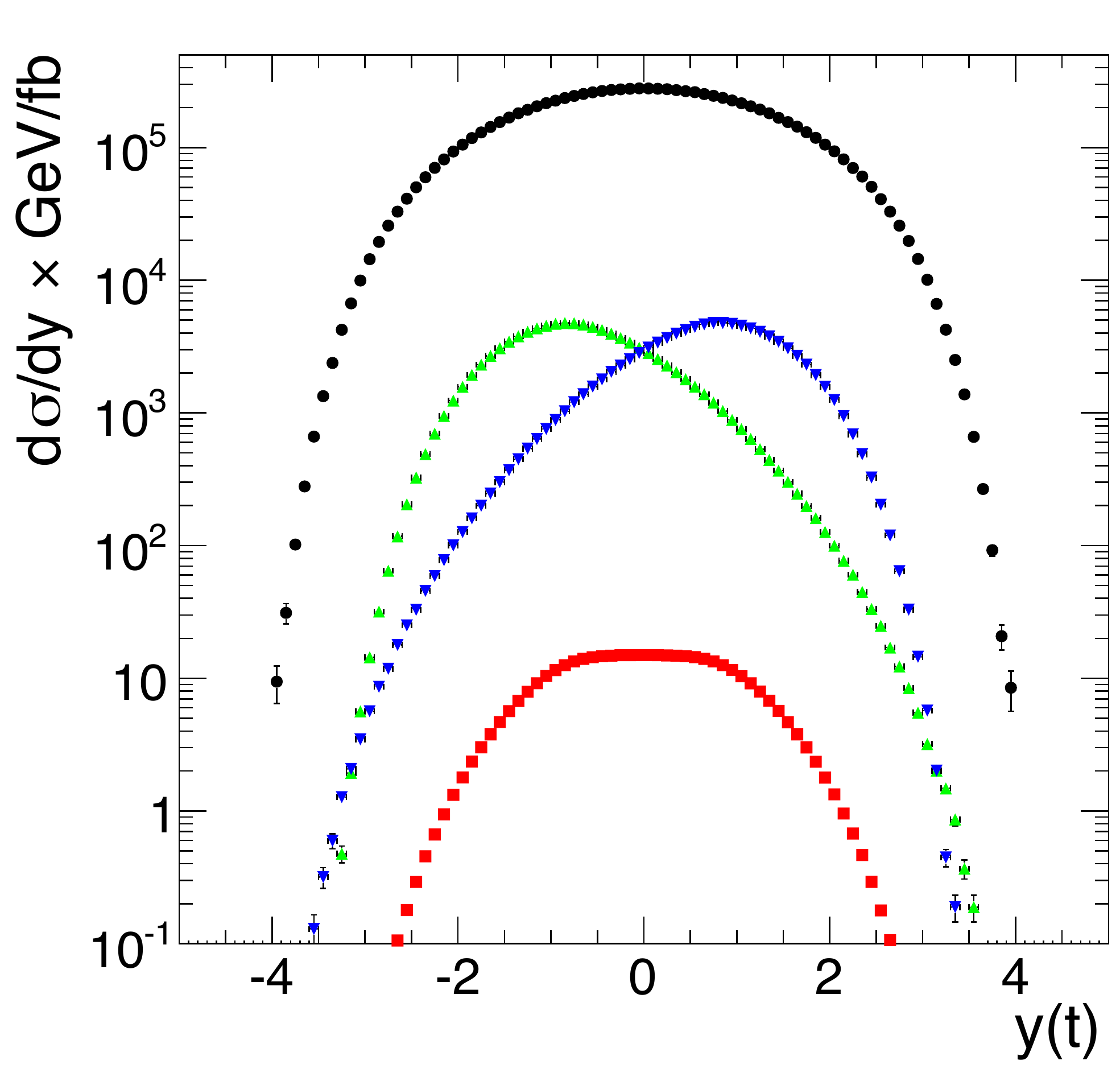}
  \end{center}
  \caption{\label{fig:ptSpectra} Impact of the Higgs induced
    sub-processes on the transverse momentum~(left) and the rapidity~(right) of the final state top
    quarks in $pp\to \ttbar$. We have used an example PDF of
    $0.01\times (1-x)\exp(-43 x^2)/x$ without any rescaling taken into account for
    sum rules, and a distribution of the intermediate virtuality
    according to the massive version of the propagator term. We
    separately show the parton induced production ('$q\bar{q}/gg$') and then the
    cross sections from the gluon-Higgs (gH) and Higgs-Higgs (HH) induced
    sub-processes. Notice that the gH and Hg sub-processes are mirror
    symmetric in the rapidity distribution, but identical in the
    $p_\perp$ distribution.}
\end{figure}

As an example of the impact of the Higgs-induced sub-processes we show
their impact on typical reconstructed top quark observables in
Figure~\ref{fig:ptSpectra}, just at parton level for the fixed-order hard
process.

\subsection{Selection of the PDF from the total cross section}\label{ss:selection}

\begin{figure}
\includegraphics[width=\textwidth]{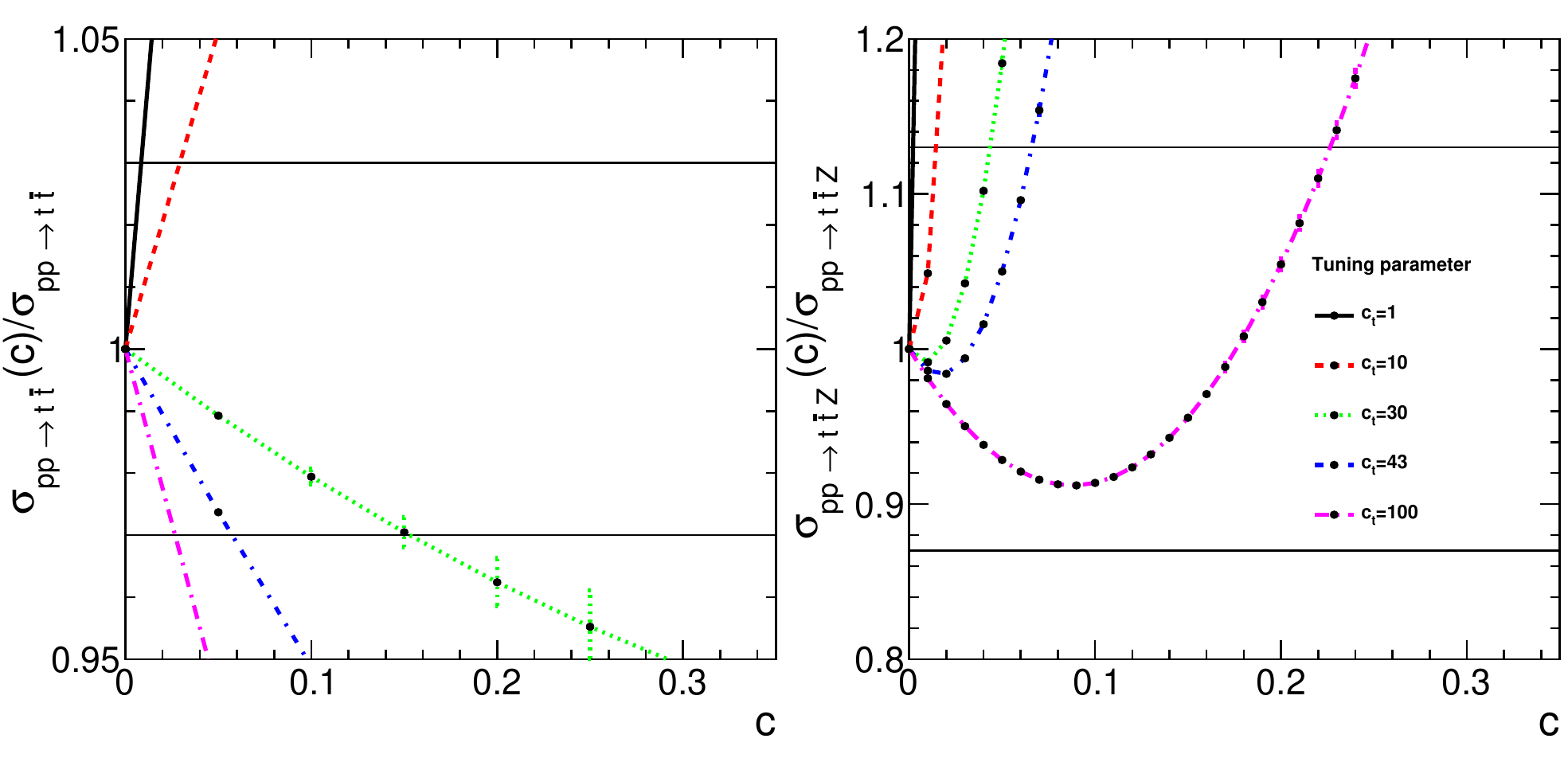}
    \caption{\label{fig:xs}The allowed band \pref{allowedband} as a function of $c$ for $\ttbar$ (left panel) and $\ttZ$ (right panel) for the PDF \pref{pdf2} for  varying values $c_t$. The horizontal lines represent $e_l^X$ and $e_u^X$.}
\end{figure}

The Higgs PDF to be used for an input into \delphes was selected by identifying the PDF which allowed for a maximal value of $c$ in \pref{totalxs} without yielding a result exceeding the experimental error margin.
Based on the achievable experimental precision~\cite{Aad:2019hzw,Defranchis:2019mvt}, we require that
\be
e_l^{\ttbar}\le \frac{\sigma_{pp\to \ttbar}(c)}{\sigma_{pp\to \ttbar}}\le e_u^{\ttbar}\label{allowedband},
\ee
\no where $e_l^{\ttbar}=0.97$, $e_u^{\ttbar}=1.03$ at 68\% CL. 
In addition, we used the process $pp\to \ttZ$ with the corresponding experimental corridor $e_l^{\ttZ}=0.87$, and $e_u^{\ttZ}=1.13$ \cite{Sirunyan:2017uzs} . 

This procedure is illustrated in Figure~\ref{fig:xs}. Depending on the tuning parameter, vastly different values for $c$ are possible. In fact, as the tuning parameter is varied, the cross section changes continuously from exceeding the experimental error bar to undershooting it. Hence, a sweet spot\footnote{As a consequence, it is possible that there are two allowed windows for the Higgs content, one ranging from zero to some fixed value, and another window between two larger values of the Higgs content. Since for none of our PDF shapes the window at larger Higgs content was compatible with the exclusion limit from the second process, we do not show this second window.} exists for which any value of $c$ is consistent. Depending on the shape of the PDF, this sweet spot for both processes can be moved closer and closer together by varying the tuning parameter, though with none of the PDF shapes we used this was possible simultaneously for both processes. Thus, both processes show a very different sensitivity to $c$, making them both complementary as constraints. Finally, the relative contributions from gH and HH initial states swap between both processes, with larger contributions from gH for the $\ttbar$ final state and from HH for $\ttZ$.

\begin{figure}
\includegraphics[width=0.5\textwidth]{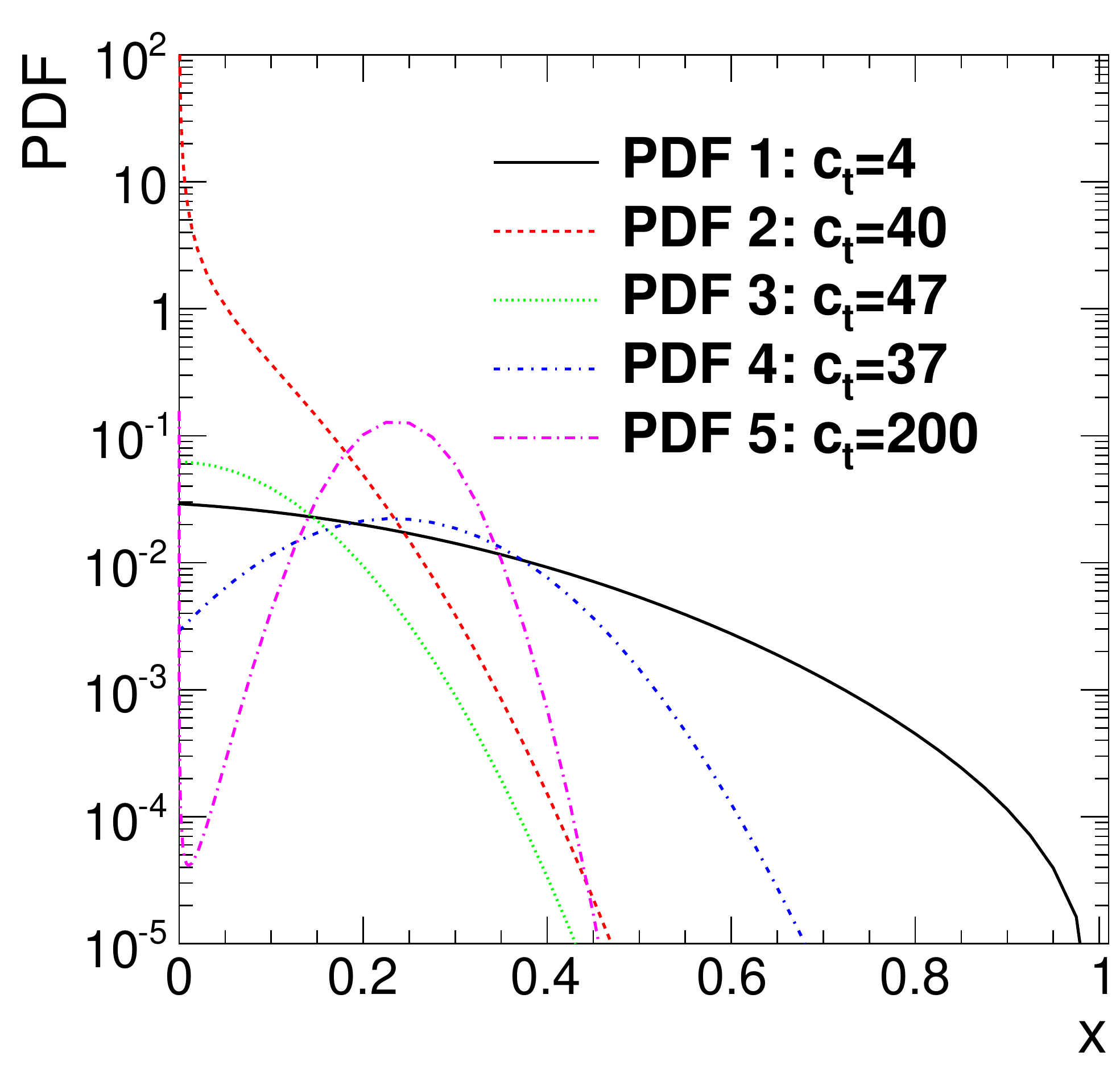}\includegraphics[width=0.5\textwidth]{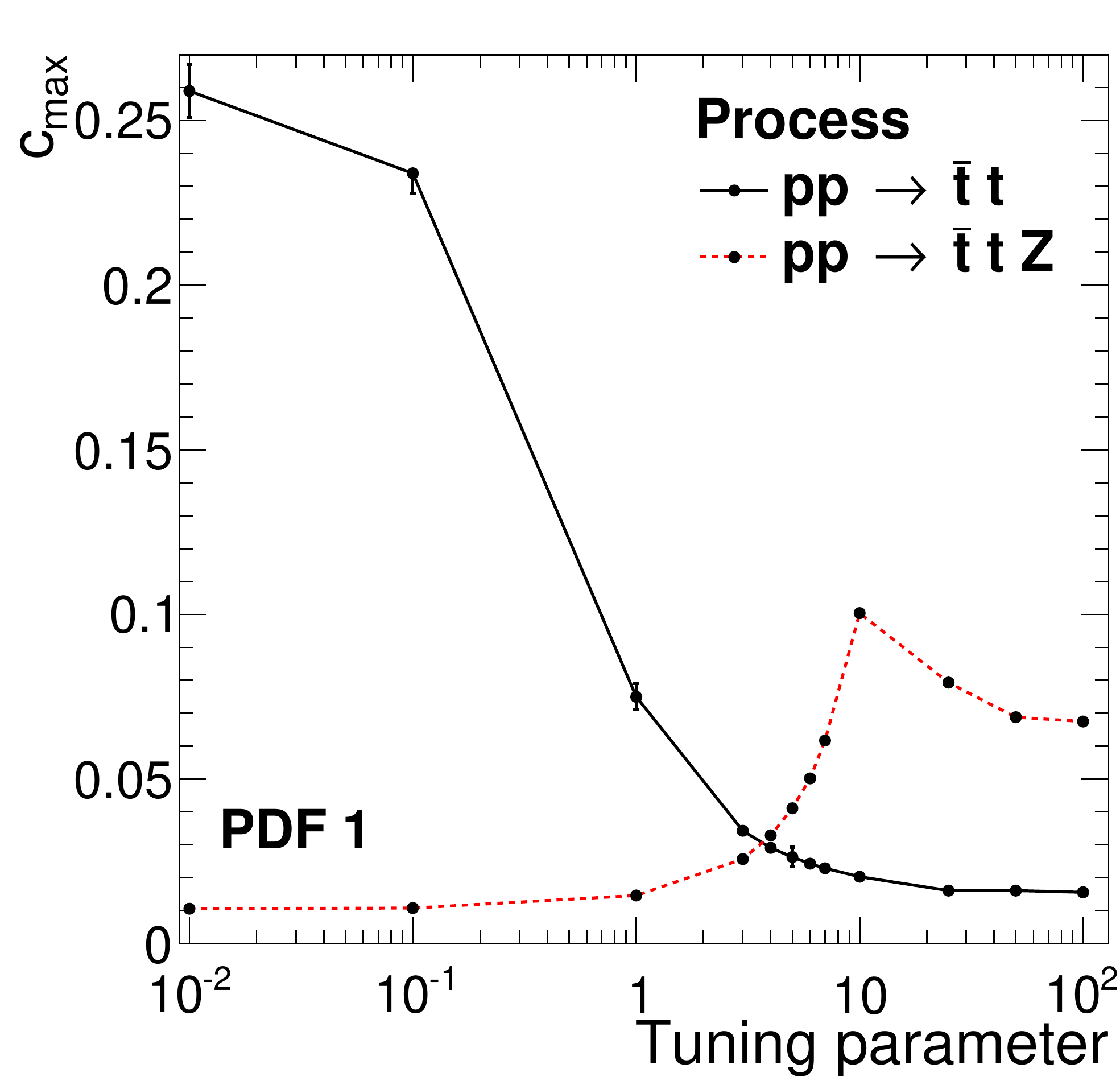}\\
\includegraphics[width=0.5\textwidth]{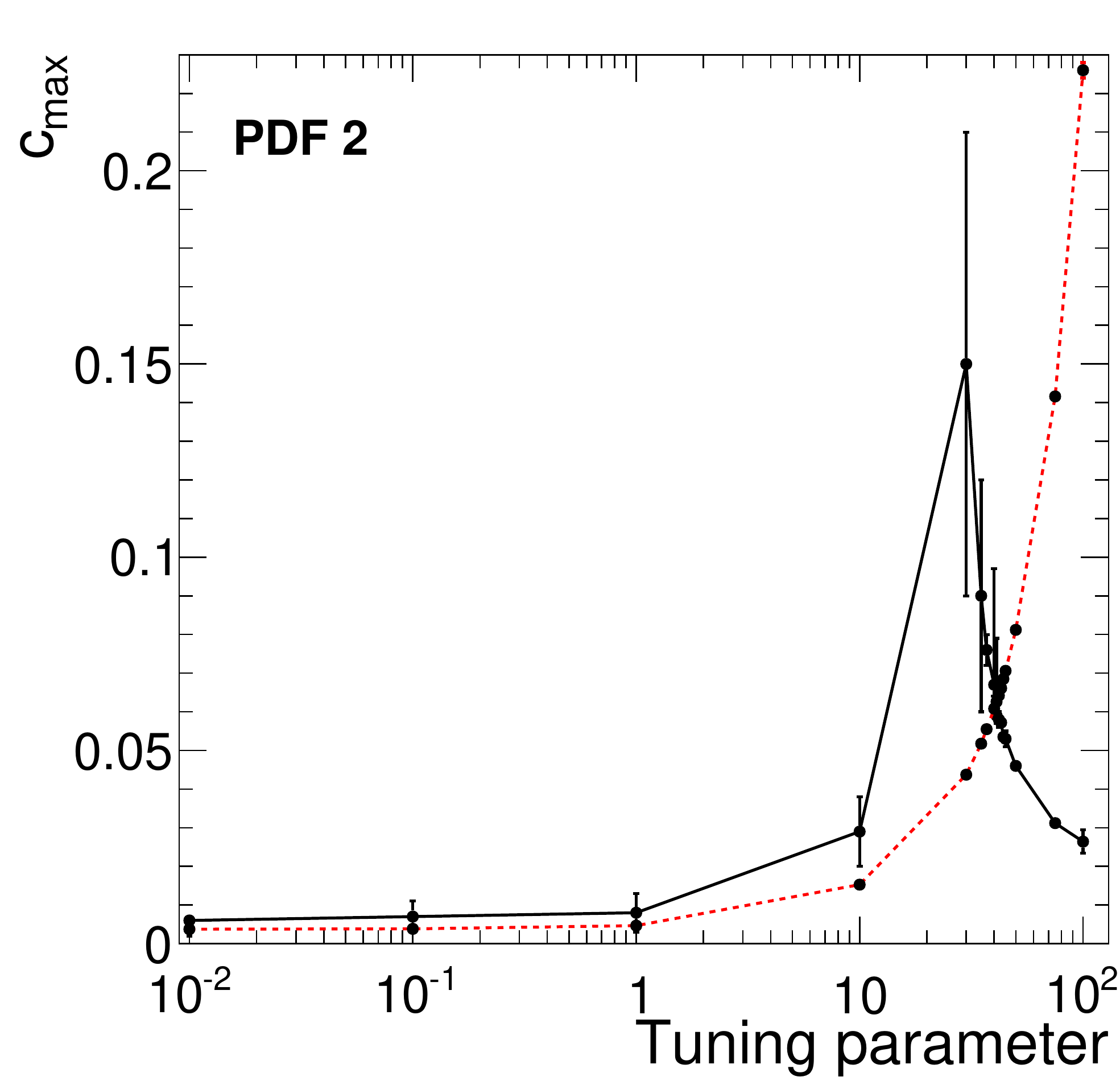}\includegraphics[width=0.5\textwidth]{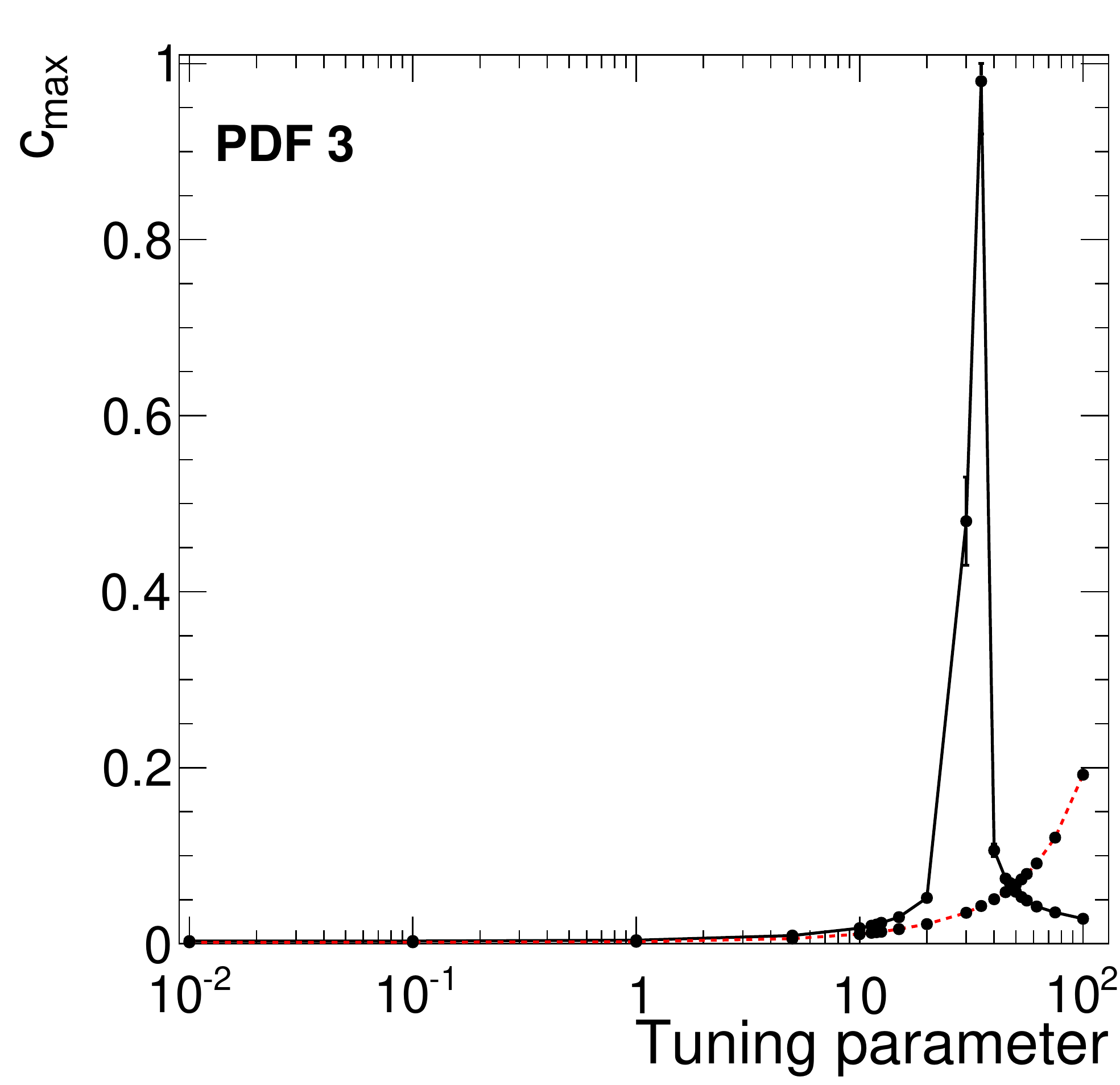}\\
\includegraphics[width=0.5\textwidth]{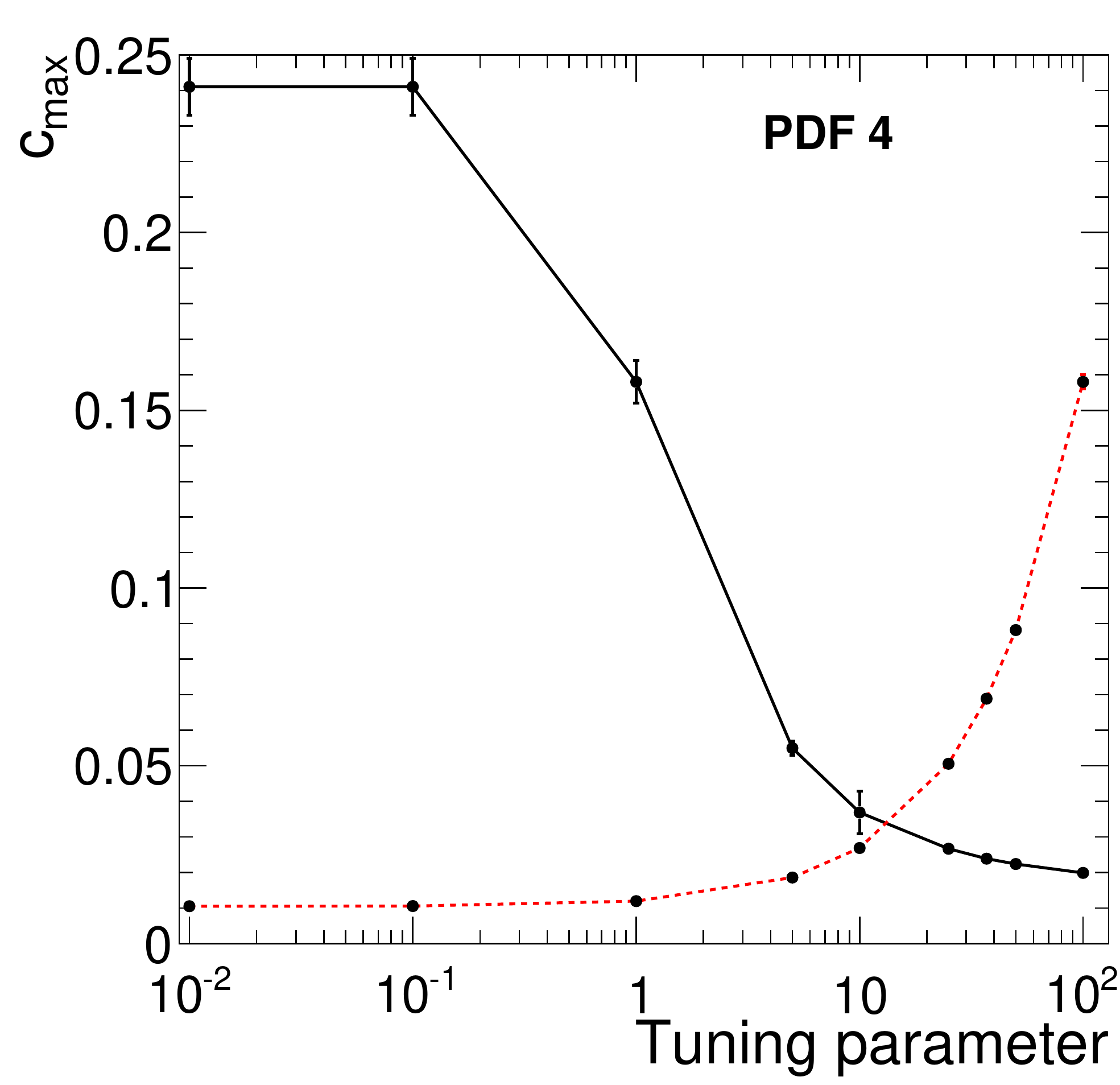}\includegraphics[width=0.5\textwidth]{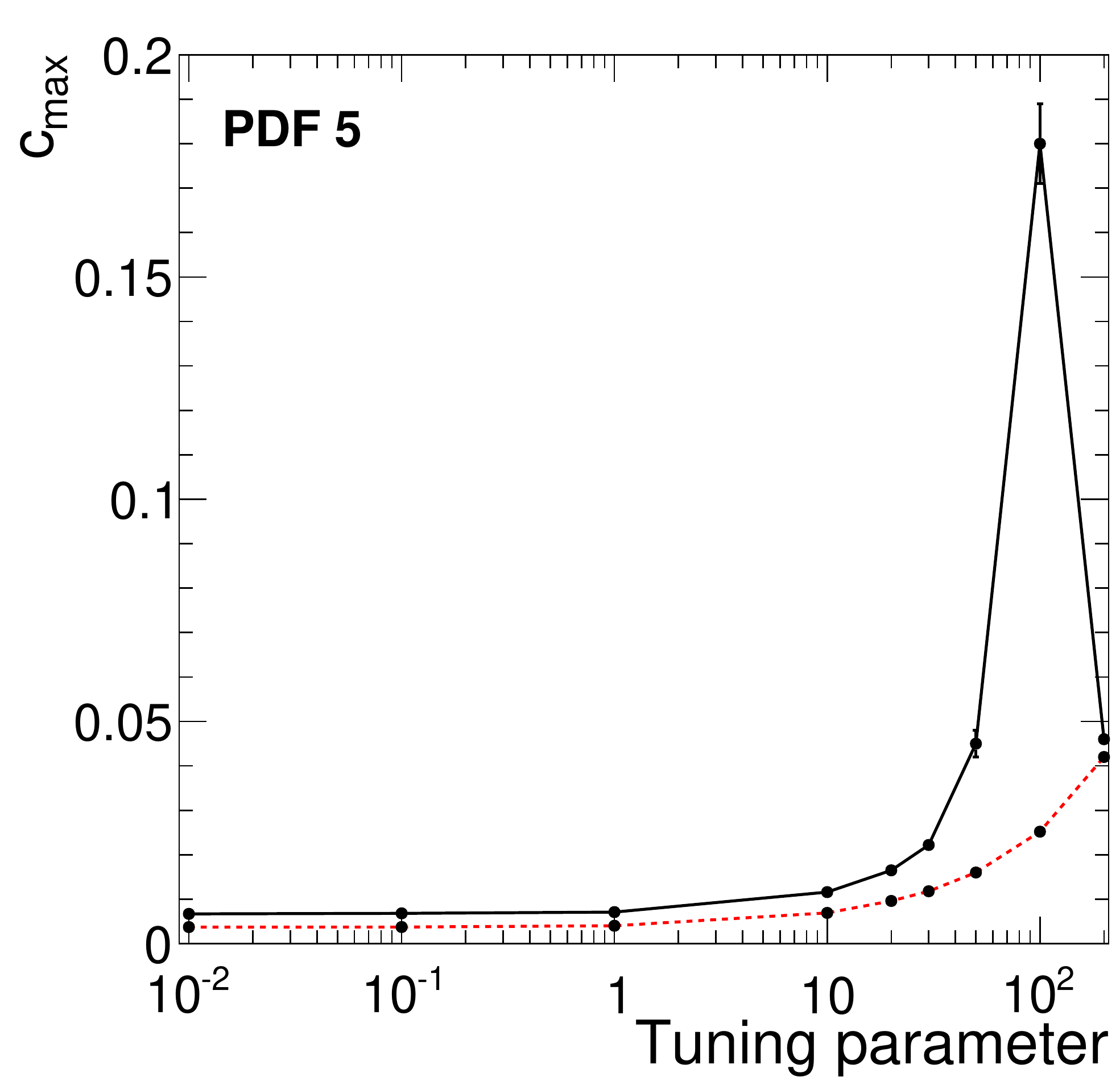}\\
    \caption{\label{fig:pdf-selection}The top-left panel shows five different version of the PDF, with parameters fixed to allow for maximal Higgs content of the proton. The other five panels show how the optimal parameter is obtained from constraints due to the total cross sections for the processes $pp\to\ttbar$ and $pp\to\ttZ$.}
\end{figure}

The resulting constraints on $c$ are shown in Figure~\ref{fig:pdf-selection} for the five different PDF versions. It is visible that both processes yield quite different constraints, marking them as complementary tests. Generically, the process $pp\to\ttZ$ is less constrained by the high-$x$ behavior, i.\ e.\ at tuning parameters where the PDFs becomes narrower around small $x$. The process $pp\to\ttbar$ is especially less sensitive to distributions extending to rather large values, but are sensitive to very narrow or extremely broad distributions. The reason is that to have a large Higgs content requires that the cross section coming from the Higgs initial states is not too different than the one from quarks and gluons. This originates from the fact that the contribution from the Higgs initial states for the $\ttbar$ final state generically decrease with increasing tuning parameter, starting at the smallest values with cross sections from different initial states much larger than the one from quarks and gluons, but being smaller at the largest value of the tuning parameter. Thus, there is a sweet spot at an intermediate value. For the $\ttZ$ final state the situation is similar, but here the HH initial state completely dominates over the gH initial state, and thus the quadratic dependency on the Higgs content drives the effect. This is due to the possibility that the $\cPZ$ is emitted in the initial state only.

Alongside this in Figure~\ref{fig:pdf-selection} also the 'optimal' PDFs are shown, i.\ e.\ the ones which allows for a maximal contribution of the Higgs to the proton. The optimal choices for the parameters yield relatively similar PDFs, with values around $5\times 10^{-2}$ for the $x$-range 0.1 to roughly 0.3, i.\ e.\ in a typical range for a valence particle. The following will now concentrate on PDF 2, which allows for the largest Higgs content, though PDF 3 yields quite similar results.

\subsection{Signatures in the final state at the partonic level}\label{ss:parton}

\begin{figure}
\includegraphics[width=0.5\textwidth]{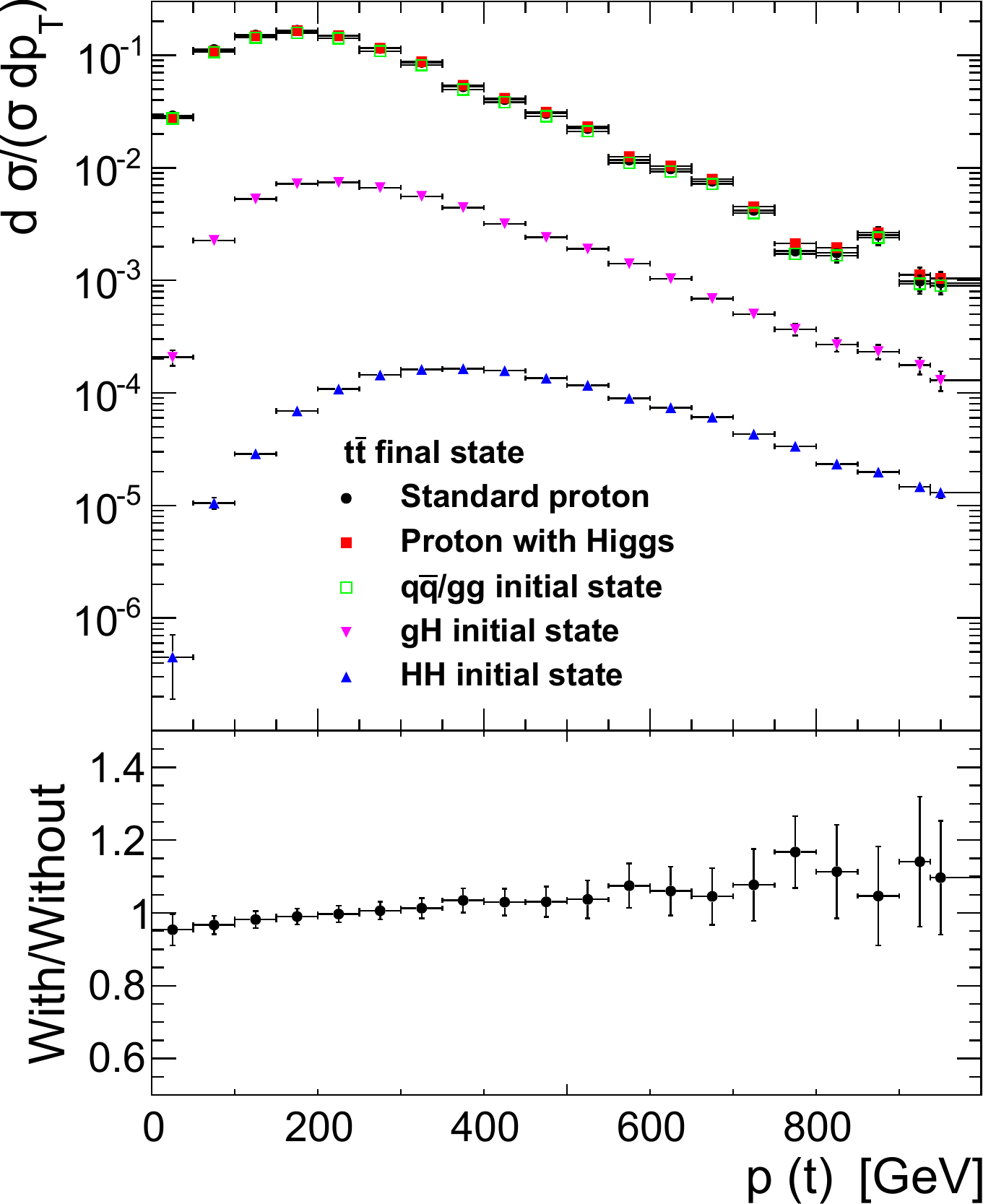}\includegraphics[width=0.5\textwidth]{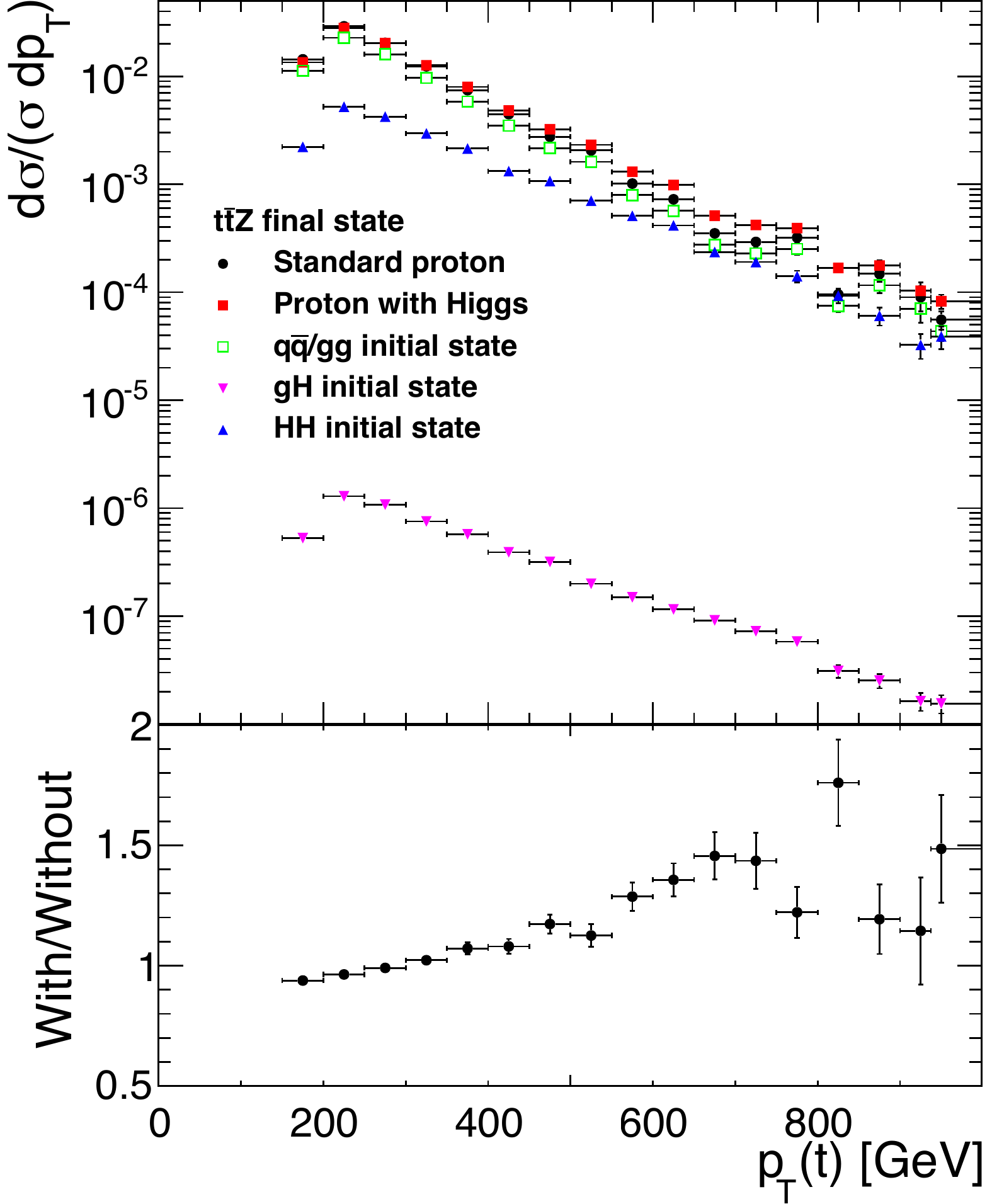}
    \caption{\label{fig:partial}The differential cross section for the $p_T$ distribution of the top for the final state $\ttbar$ (left panel) and $\ttZ$ (right panel). The top panels show the comparison between the initial state without and with valence Higgs. The latter is shown also broken down into its different contribution of partonic initial states. The bottom panels show the ratio of the differential cross sections for both initial states. This is PDF 2 with tuning parameter $c_t=40$ and Higgs content $c=0.0608$. Errors are Herwig errors for 10000 events.}
\end{figure}

As will become clear in section \ref{s:cms}, the differential cross sections at the detector level will further reduce the maximum Higgs content of the proton substantially. It is worthwhile to investigate the origin of this effect at the level of the not-hadronized final state. While many different partial cross sections will be used later in section \ref{s:cms}, there is a generic behavior. This is illustrated in Figure~\ref{fig:partial}, showing the differential cross section for the $p_T$ spectrum of the top.

It is visible that the spectrum of the tops gets harder when the initial state contains a Higgs in the initial state. Especially with two Higgs in the initial state the spectrum is even harder than with one Higgs in the initial state. Hence, no matter how large the Higgs content, at sufficiently high transverse momentum~($p_T$) of the top eventually a deviation will arise. Thus, as experiments grow more sensitive at large values of $p_T$, the Higgs content is further reduced, if no enhancement is observed. In fact, the corresponding experimentally observed spectrum \cite{Sirunyan:2019zvx} currently shows even a suppression compared to the expectation from the case without valence Higgs, creating an even stronger restriction of the Higgs content.

This feature seems to be essentially generic within the set of PDFs we have investigated. It may be that different shapes, $Q^2$ evolution, or a complete refit of all PDFs may change this behavior. If not, then an unambiguous signal for the Higgs content of the proton would be a hardening of the top spectrum in this process.

Note that a similar enhancement of top production with increasing energy in the presence of initial state Higgs has also been observed in the case of lepton collisions in \cite{Egger:2017tkd}, with a much simpler Higgs PDF. With bottoms or muons in the final state no such enhancement has been observed, suggesting that kinematic effects may be relevant.

\section{Experimental constraints}\label{s:cms}

In order to estimate current bounds from experiment we perform event reconstruction with \delphes~\cite{deFavereau:2013fsa}, where we use the CMS reconstruction efficiency parametrisation for the LHC Run II. 
The $\ttbar$ process in the single lepton final state allows a relatively pure signal selection and eschews the lower branching ratio of the \cPW\;bosons in the dilepton final states.
Jets are reconstructed with the {\texttt FastJet} package~\cite{Cacciari:2011ma} and with the anti-$k_T$ algorithm~\cite{Cacciari:2008gp} with a cone size of $R=0.4$.
As described before, we simulate the $\ttbar$ process including the diagrams with an initial state Higgs boson with Herwig.
Besides this process, we also generate the main backgrounds in the leptonic final states in order to achieve a realistic background prediction.
Important backgrounds to the $\ttbar$ process include the $\wjets$ process where we include up to three extra partons, and $\tW$ production. 
We also simulate other small contributions from the $\ttZ$, $\ttW$, $\tZ$, $\tWZ$, $\WZ$ and $\Zg$ processes whose yield are mostly negligible.
The backgrounds are generated at the parton level events at LO using \MGvATNLO v2.3.3~\cite{Alwall:2014hca}, and decayed using {\tt MadSpin}~\cite{Artoisenet:2012st, Frixione:2007zp}.
Parton showering and hadronisation for the backgrounds are performed with \PYTHIA 8.2~\cite{Sjostrand:2007gs,Sjostrand:2014zea}.
The $\wjets$ sample is simulated at leading order in perturbative QCD, while all other background processes are simulated at next-to-leading order with \MGvATNLO.
The statistical uncertainties on the background contributions are negligible in all cases.
For validation purposes, we simulate the $\ttbar$ process excluding the valence Higgs distribution following the procedure of background simulation.
We find very good agreement of all relevant kinematic distributions between the Herwig~7 and the \MGvATNLO simulation. The generated samples are summarized in Table~\ref{tab:samples}.

After event reconstruction, we require exactly one lepton (e or $\mu$) satisfying a threshold of $p_T(l)>30$~\GeV and $|\eta(l)|<2.4$ and a tight isolation criterion.
Furthermore, at least 4 jets with $p_T(j)>30$~GeV and $|\eta(j)|<2.5$, where at least one of the jets has to be identified as a b-tag jet according to the \delphes specification, are required. 
We remove reconstructed leptons within a cone of $\Delta R < 0.3$ of any reconstructed jet satisfying $p_T>30$~\GeV. 
These object definitions and the event selection criteria are mildly tuned to reproduce simulated distributions in Ref.~\cite{Sirunyan:2018wem}.

\begin{figure}
    \includegraphics[width=0.5\textwidth]{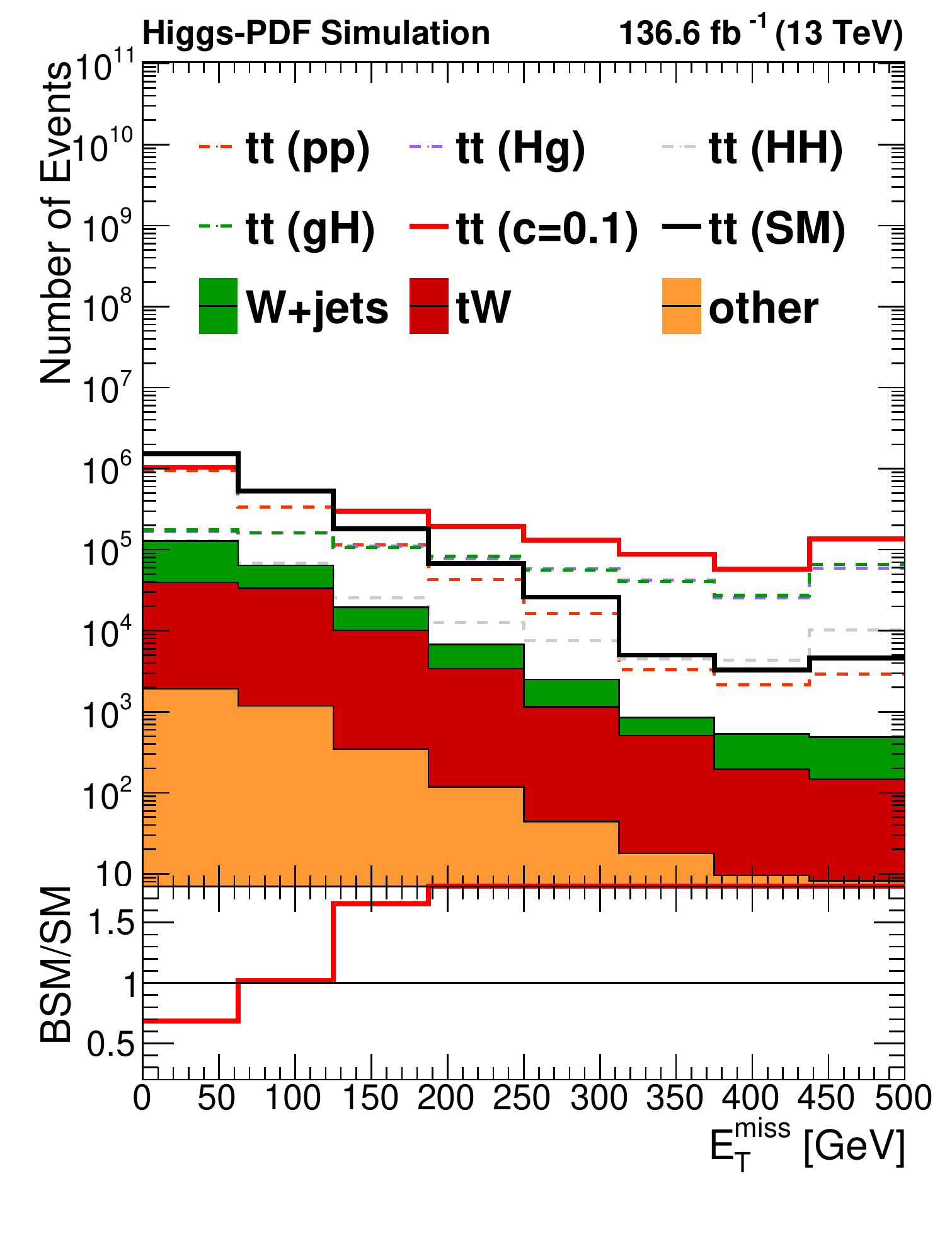}
    \includegraphics[width=0.5\textwidth]{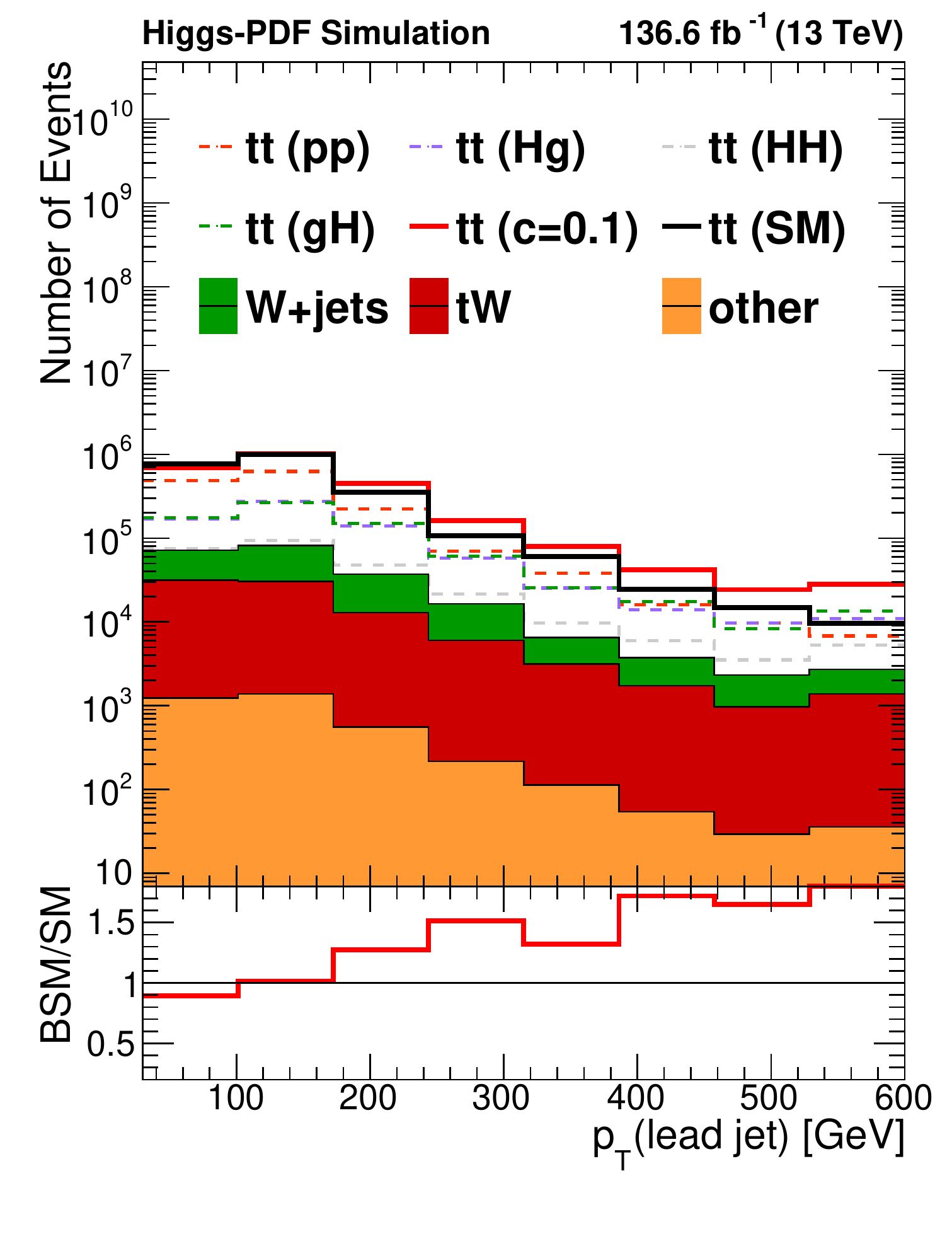}
    \caption{Simulated distribution of \MET~(left) and the $p_T$ of the leading jet~(right) for Higgs PDF 2 in the event selection described in the text. }
\label{fig:reco}
\end{figure}

In order to assess the sensitivity of a differential cross section measurement to the Higgs valence contributions, we tested several observables for their discriminative power. 
The most promising candidates were found to be simple measures of the total momentum scale of the event. 
Angular observables and dimensionless ratios are less discriminative. 
In Figure~\ref{fig:reco} we show the spectrum of missing energy~(\MET) and the $p_T$ of the leading jet in the event.
We defer refinements of such observables to later work and for now construct a simple rectangular grid of signal regions in terms of these two observables.
The thresholds for $\MET$ are chosen as 0, 100, 200, and 300~\GeV and for the leading jet $p_T$ as 0, 100, 200, and 400~\GeV. 
This coarse binning ensures that each of the resulting 16 signal regions is still populated by the signal and background simulation and, moreover, that the precise values
of the bin boundaries have no significant impact on the result. Besides that, no optimisation is attempted.
Effects of the Higgs PDF on processes other than $\ttbar$ are neglected.

The predicted yields in these regions are estimated for the 136.6~fb${}^{-1}$ LHC Run-II scenario at $\sqrt{s}=13$~\TeV.
We consider a ball-park mock up of experimental uncertainties that attempt to reflect the uncertainties in a future experimental measurement.
The sources of systematic uncertainties we consider are the jet energy scale ($\leq 5\%$), the rate of failure to identify a b-jet~($\leq 2\%$), the b-tagging rate of light flavor jets~($\leq 1\%$), the lepton identification~($\leq 1\%$) and the luminosity measurement~(2.6\%). 
The number in parenthesis are typical upper bounds for the uncertainties in signal and background yields and do not strongly vary across the signal regions.
They are obtained by a suitable reweighting of the simulated response and tagging efficiencies.
Again, the results do not depend significantly on the details of this scenario.

These uncertainties are associated with log-normal nuisance parameters $\theta$ which are used to construct a likelihood function $L(\theta)$ where $\theta$ labels the set of nuisance parameters.
We perform a profiled maximum likelihood fit of $L(\theta)$ and consider $q(c)=-2\log(L(\hat{\theta}_c)/L(\hat{\theta}_{\textrm{0}}))$, where $\hat{\theta}_c$ and
$\hat{\theta}_\textrm{0}$ are the set of nuisance parameters maximising the likelihood function at a fixed value of c and for $c=0$, respectively.

\begin{table}
\caption{Simulated background processes and event counts. Here, $\lep^\pm=e^\pm,\mu^\pm,\tau^\pm$ and $\nu_l=\nu_e,\nu_\mu,\nu_\tau$. 
The $\ttbar$ process is simulated to validate the Herwig~7 signal simulation. }
\label{tab:samples}
\begin{center}
\begin{tabular}{c|c|c|r}
\hline 
process & & $N_{\textrm{events}}$ & cross section\\ 
\hline
$\ttbar$ & $\pp\rightarrow\ttbar$ & $10^7$ & 831.76 pb  \\ 
$\wjets$ & $\pp\rightarrow\cPW+\textrm{jets},\,\cPW\rightarrow \lep\nu_l$ & $10^6$ &  61.5 nb  \\ 
$\tW$ & $\pp\rightarrow \cPqt\,\Wmlep + \pp\rightarrow \cPaqt\,\Wplep$ & $10^6$ & 19.55 pb  \\ 
$\tZ$ & $\pp\rightarrow \cPqt\,\Zoff\,\jet + \pp\rightarrow \cPaqt\,\Zoff\,\jet$ & $10^6$ & 0.0758 pb  \\ 
$\ttZ$ & $\pp\rightarrow\ttbar\,\Zoff$ & $10^6$ & 0.0915 pb  \\ 
$\ttW$ & $\pp\rightarrow\ttbar\,\Wmlep + \pp\rightarrow\ttbar\,\Wplep$ & $10^6$ & 0.2043 pb  \\ 
$\tWZ$ & $\pp\rightarrow \cPqt\,\cPW\,\Zoff + \pp\rightarrow \cPaqt\,\cPW\,\Zoff$ & $10^6$ & 0.01123 pb  \\ 
$\WZ$ & $\pp\rightarrow \Wplep\,\Zoff + \pp\rightarrow \Wmlep\,\Zoff$ & $10^6$ & 4.666 pb  \\ 
$\Zg$ & $\pp\rightarrow \Zoff\,\gamma + \pp\rightarrow \Zoff\,\gamma\,\jet$ & $10^6$ & 131.3 pb  \\ 
\hline
\end{tabular}
\end{center}
\end{table} 

\begin{figure}
    \includegraphics[width=0.5\textwidth]{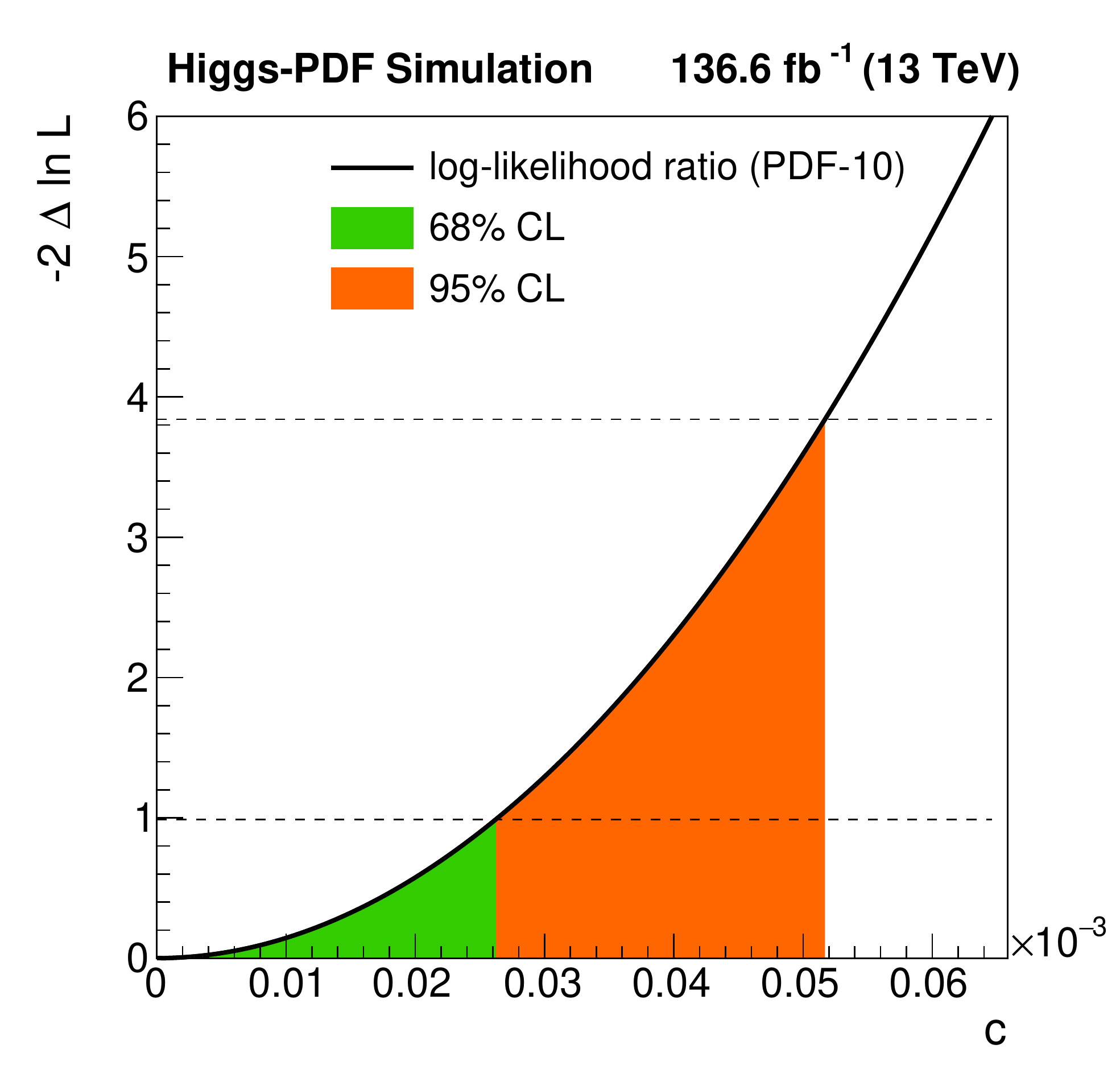}
    \includegraphics[width=0.5\textwidth]{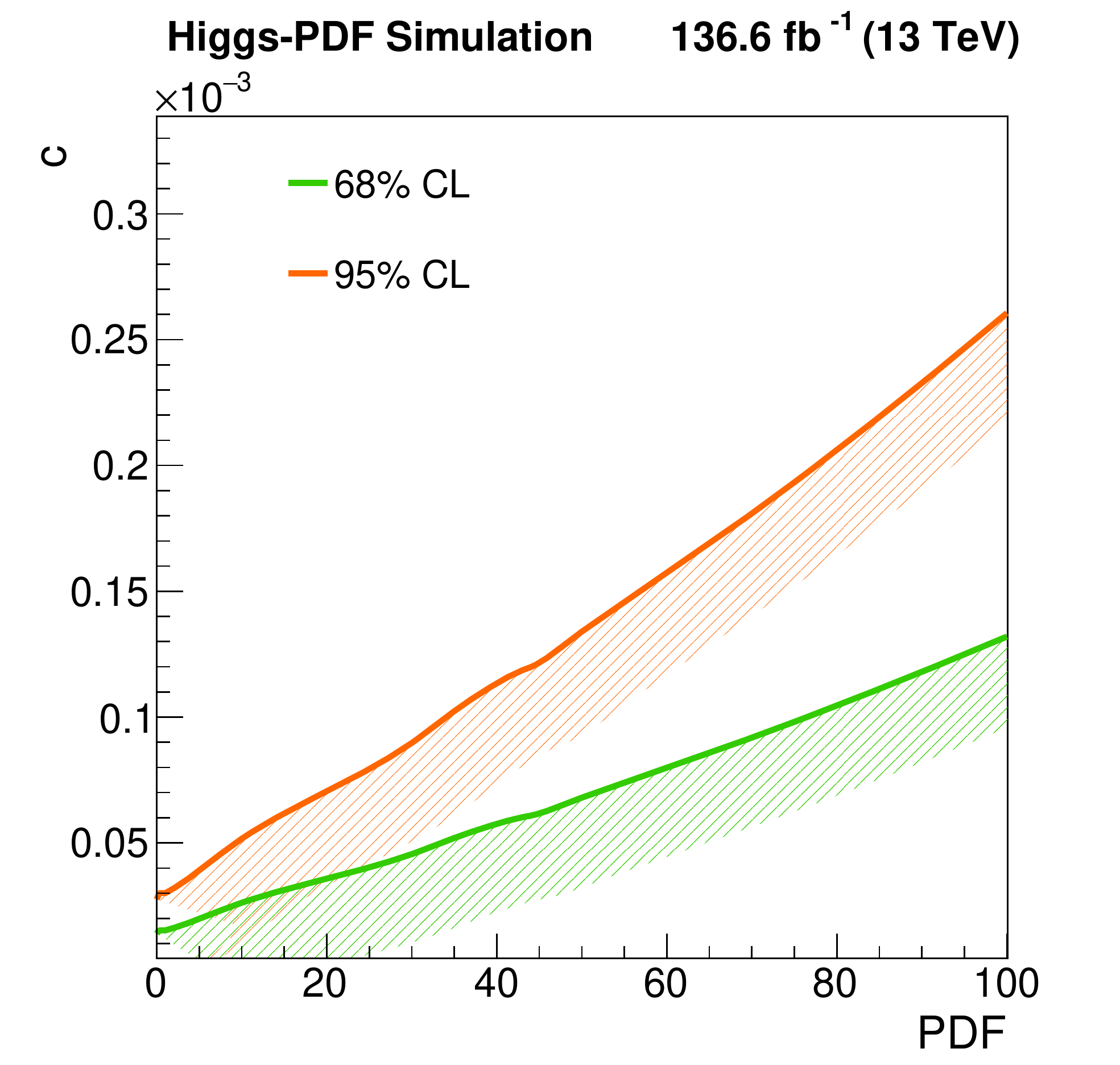}
    \caption{The left-hand side shows the upper limits at 68\% and 95\% confidence level as a function of the tuning parameter from the analysis with \delphes for Higgs PDF 2 at $c_t=35$. The right-panel shows the resulting optimal Higgs PDF overall size as a function of the Higgs PDF 2 tuning value.}
\label{fig:limits}
\end{figure}

\section{Results}\label{s:results}

The procedure leads to upper limits of the Higgs content of the proton as a function of the tuning parameters. 
As an example, Figure~\ref{fig:limits}~(left) shows $q(c)$ for a tuning parameter of 35 with the 68\% and the 95\% confidence levels (CL) indicated. 
In Figure~\ref{fig:limits}~(right), we show the 68\% and 95\% CL as a function of the tuning parameter.
Values of c in excess of $2.5\times 10^{-4}$ are excluded at 95\% CL for the considered tuning parameter range, with the strongest constraints from the top quarks at high transverse momentum.
The importance of highly energetic tails is consistent with the findings in Section~\ref{ss:parton}, 
and in fact suggests that constraints at future facilities such as the HL-LHC scale favorably with luminosity. 
Because the contribution from the HH initial state is proportional to $c^2$, it is negligible at or below the obtained expected limits. 
While no attempt was made to obtain observed limits, it should be noted that there currently is no significant excess in the tails of kinematic spectra of top quarks or the $\ttbar$ decay products~\cite{Sirunyan:2018wem,Sirunyan:2019zvx}, and
therefore we expect that future studies will find consistent observed limits.
We defer the study of the indirect effect of a non-zero proton Higgs PDF to the precisely measured final states to later work.

\section{Discussion}\label{s:discussion}

\begin{figure}
\begin{center}
\includegraphics[width=0.75\textwidth]{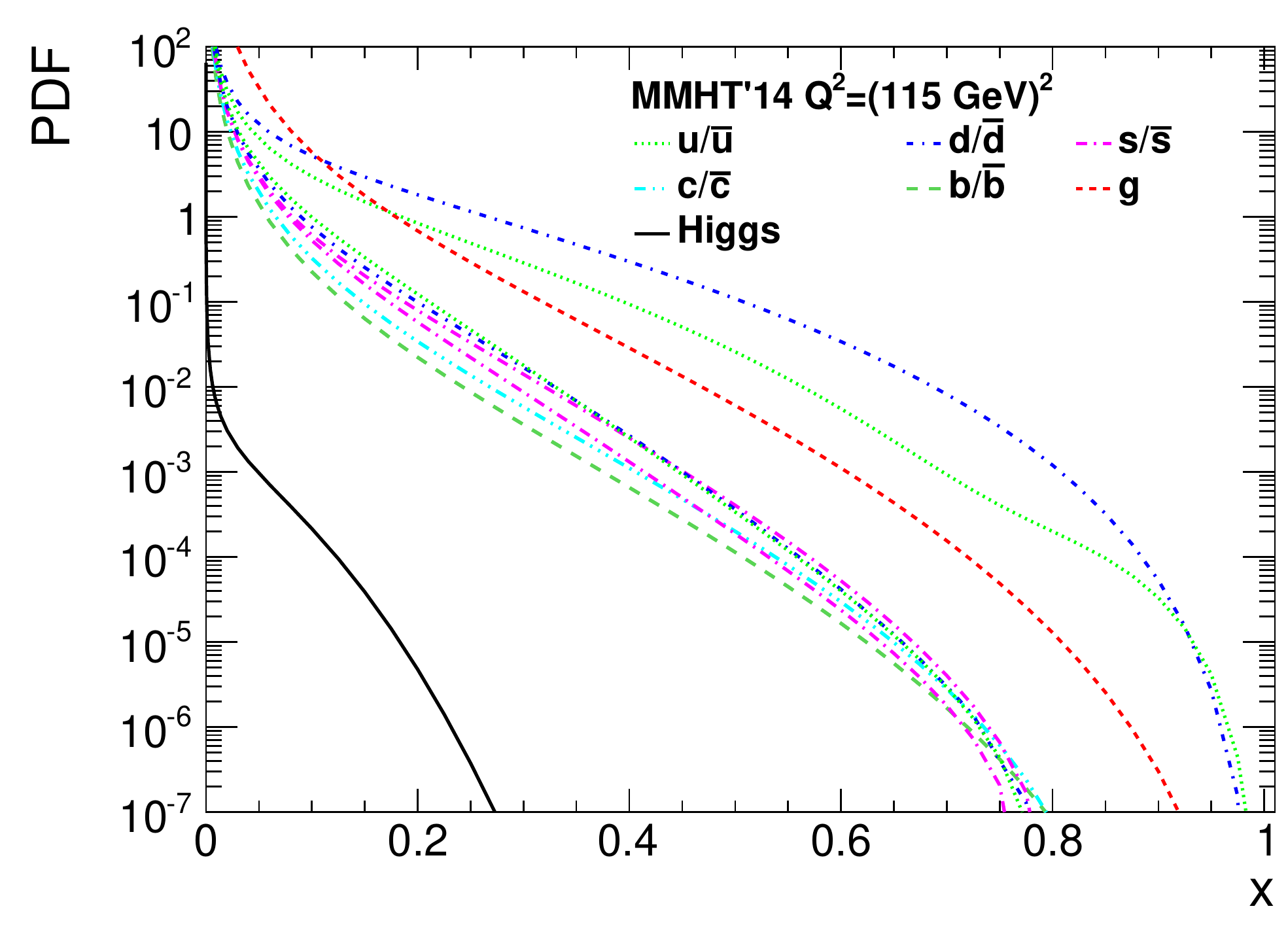}
\end{center}
    \caption{\label{fig:finalpdf}The Higgs PDF 2 corresponding to the upper limit at 95\% confidence level for a tuning parameter $c_t=100$ in comparison to the remaining PDFs for the MMHT'14 set at a $Q^2$ corresponding approximately to the Higgs mass.}
\end{figure}

We have obtained the first expected constraints on a Higgs valence contribution of the proton using simple functional parametrisations of the valence Higgs PDF demanded by field theory. 
The results show that the $\ttbar$ and $\ttbar\PZ$ processes constrain a valence Higgs contribution and the
allowed contribution is certainly substantially smaller than the next largest PDF for the bottom quark~(Figure~\ref{fig:finalpdf}). 
This is expected owing to the comparably large mass of the Higgs boson. 
Future experiments at hadron colliders (HL-LHC, FCC-hh) will be sensitive to the valence Higgs contributions in the tails of kinematic distributions.
On the theoretical side, the next logical step is a global PDF fit including a Higgs PDF and its $Q^2$ evolution. 

While the results here certainly encourage the compatibility of a Higgs PDF with the data, this opens two important questions: If a global refit does not yield an essentially vanishing Higgs PDF, in contrast to the results here, then it appears there is an ambiguity in fitting the data with PDFs. It would imply that an additional PDF could have been covered by the other PDFs so far. That appears somewhat unlikely given the considerations in \cite{Carrazza:2019sec}. A prerequisite to such a study is to formalize factorization theorems involving electroweak particles in the initial state, something which we have only made a phenomenologically reasonable assumption on. Such a treatment should clarify how the actual amplitudes can be evaluated and how NLO QCD corrections can be included for the Higgs-induced processes.

On the other hand, taking the perspective that the field theoretical motivation is solid for the presence of the Higgs PDF, this implies that if the data cannot be consistently fitted with the additional Higgs PDF, this could already imply new physics. After all, the additional Higgs component adds further leading-order interaction channels, which had been absent so far. An example is the aforementioned too soft top spectrum \cite{Sirunyan:2019zvx}. The additional Higgs component makes it even harder to make this consistent with the SM, and thus enlarges the discrepancy of the measurement to the SM.

In addition, by choosing suitable kinematic cuts to prefer initial state Higgs, this may also extend the reach for new physics searches, if they couple dominantly to the Higgs, e.\ g.\ dark matter through the Higgs portal.

\section*{Acknowledgments}

SP is grateful to the other Herwig members for useful discussions and
in particular Peter Richardson for exchange and support on several
bug fixes appearing in the course of this work.  This work has been
supported in part by the European Union’s Horizon 2020 research and
innovation programme as part of the Marie Skłodowska-Curie Innovative
Training Network MCnetITN3 (grant agreement no. 722104). SP also
acknowledges partial support by the COST actions CA16201
``PARTICLEFACE'' and CA16108 ``VBSCAN''.
 
\bibliographystyle{bibstyle}
\bibliography{bib}

\begin{thebibliography}{10}

\bibitem{Bohm:2001yx}
M.~B\"ohm, A.~Denner, and H.~Joos,
\newblock {\em {Gauge theories of the strong and electroweak interaction}}
  (Teubner, Stuttgart, 2001).

\bibitem{'tHooft:1979bj}
G.~'t~Hooft,
\newblock NATO Adv.Study Inst.Ser.B Phys. {\bf 59}, 101 (1980).

\bibitem{Gribov:1977wm}
V.~N. Gribov,
\newblock Nucl. Phys. {\bf B139}, 1 (1978).

\bibitem{Singer:1978dk}
I.~M. Singer,
\newblock Commun. Math. Phys. {\bf 60}, 7 (1978).

\bibitem{Fujikawa:1982ss}
K.~Fujikawa,
\newblock Nucl. Phys. {\bf B223}, 218 (1983).

\bibitem{Banks:1979fi}
T.~Banks and E.~Rabinovici,
\newblock Nucl.Phys. {\bf B160}, 349 (1979).

\bibitem{Frohlich:1980gj}
J.~Fr\"ohlich, G.~Morchio, and F.~Strocchi,
\newblock Phys.Lett. {\bf B97}, 249 (1980).

\bibitem{Frohlich:1981yi}
J.~Fr\"ohlich, G.~Morchio, and F.~Strocchi,
\newblock Nucl.Phys. {\bf B190}, 553 (1981).

\bibitem{Maas:2015gma}
A.~Maas,
\newblock Mod. Phys. Lett. {\bf A30}, 1550135 (2015), 1502.02421.

\bibitem{Maas:2017xzh}
A.~Maas, R.~Sondenheimer, and P.~T\"orek,
\newblock Annals of Physics {\bf 402}, 18 (2019), 1709.07477.

\bibitem{Sondenheimer:2019idq}
R.~Sondenheimer,
\newblock (2019), 1912.08680.

\bibitem{Maas:2016ngo}
A.~Maas and P.~T\"orek,
\newblock Phys. Rev. {\bf D95}, 014501 (2017), 1607.05860.

\bibitem{Maas:2018xxu}
A.~Maas and P.~T\"orek,
\newblock Annals Phys. {\bf 397}, 303 (2018), 1804.04453.

\bibitem{Maas:2017wzi}
A.~Maas,
\newblock Progress in Particle and Nuclear Physics {\bf 106}, 132 (2019),
  1712.04721.

\bibitem{Maas:2012tj}
A.~Maas,
\newblock Mod.Phys.Lett. {\bf A28}, 1350103 (2013), 1205.6625.

\bibitem{Maas:2013aia}
A.~Maas and T.~Mufti,
\newblock JHEP {\bf 1404}, 006 (2014), 1312.4873.

\bibitem{Egger:2017tkd}
L.~Egger, A.~Maas, and R.~Sondenheimer,
\newblock Mod. Phys. Lett. {\bf A32}, 1750212 (2017), 1701.02881.

\bibitem{Maas:2018ska}
A.~Maas, S.~Raubitzek, and P.~Törek,
\newblock Phys. Rev. {\bf D99}, 074509 (2019), 1811.03395.

\bibitem{Bauer:2017isx}
C.~W. Bauer, N.~Ferland, and B.~R. Webber,
\newblock JHEP {\bf 08}, 036 (2017), 1703.08562.

\bibitem{Bauer:2018xag}
C.~W. Bauer, D.~Provasoli, and B.~R. Webber,
\newblock JHEP {\bf 11}, 030 (2018), 1806.10157.

\bibitem{Bauer:2018arx}
C.~W. Bauer and B.~R. Webber,
\newblock JHEP {\bf 03}, 013 (2019), 1808.08831.

\bibitem{Bellm:2015jjp}
J.~Bellm {\em et~al.},
\newblock Eur. Phys. J. {\bf C76}, 196 (2016), 1512.01178.

\bibitem{deFavereau:2013fsa}
DELPHES 3, J.~de~Favereau {\em et~al.},
\newblock JHEP {\bf 02}, 057 (2014), 1307.6346.

\bibitem{Chatrchyan:2008aa}
CMS, S.~Chatrchyan {\em et~al.},
\newblock JINST {\bf 3}, S08004 (2008).

\bibitem{Maas:2019dwd}
A.~Maas {\em et~al.},
\newblock {Probing standard-model Higgs substructures using tops and weak gauge
  bosons},
\newblock in {\em {2019 European Physical Society Conference on High Energy
  Physics (EPS-HEP2019) Ghent, Belgium, July 10-17, 2019}}, 2019, 1910.14316.

\bibitem{Bahr:2008pv}
M.~Bahr {\em et~al.},
\newblock Eur. Phys. J. {\bf C58}, 639 (2008), 0803.0883.

\bibitem{Bellm:2017bvx}
J.~Bellm {\em et~al.},
\newblock (2017), 1705.06919.

\bibitem{Platzer:2011bc}
S.~Platzer and S.~Gieseke,
\newblock Eur. Phys. J. {\bf C72}, 2187 (2012), 1109.6256.

\bibitem{Gieseke:2003rz}
S.~Gieseke, P.~Stephens, and B.~Webber,
\newblock JHEP {\bf 12}, 045 (2003), hep-ph/0310083.

\bibitem{Alwall:2014hca}
J.~Alwall {\em et~al.},
\newblock JHEP {\bf 07}, 079 (2014), 1405.0301.

\bibitem{Sjodahl:2014opa}
M.~Sjodahl,
\newblock Eur. Phys. J. {\bf C75}, 236 (2015), 1412.3967.

\bibitem{Buckley:2014ana}
A.~Buckley {\em et~al.},
\newblock Eur. Phys. J. {\bf C75}, 132 (2015), 1412.7420.

\bibitem{Aad:2019hzw}
ATLAS, G.~Aad {\em et~al.},
\newblock (2019), 1910.08819.

\bibitem{Defranchis:2019mvt}
ATLAS, CMS, M.~M. Defranchis,
\newblock (2019), 1901.10898.

\bibitem{Sirunyan:2017uzs}
CMS, A.~M. Sirunyan {\em et~al.},
\newblock JHEP {\bf 08}, 011 (2018), 1711.02547.

\bibitem{Sirunyan:2019zvx}
CMS, A.~M. Sirunyan {\em et~al.},
\newblock Submitted to: Eur. Phys. J.  (2019), 1904.05237.

\bibitem{Cacciari:2011ma}
M.~Cacciari, G.~P. Salam, and G.~Soyez,
\newblock Eur.Phys.J. {\bf C72}, 1896 (2012), 1111.6097.

\bibitem{Cacciari:2008gp}
M.~Cacciari, G.~P. Salam, and G.~Soyez,
\newblock JHEP {\bf 0804}, 063 (2008), 0802.1189.

\bibitem{Artoisenet:2012st}
P.~Artoisenet, R.~Frederix, O.~Mattelaer, and R.~Rietkerk,
\newblock JHEP {\bf 03}, 015 (2013), 1212.3460.

\bibitem{Frixione:2007zp}
S.~Frixione, E.~Laenen, P.~Motylinski, and B.~R. Webber,
\newblock JHEP {\bf 04}, 081 (2007), hep-ph/0702198.

\bibitem{Sjostrand:2007gs}
T.~Sj{\"o}strand, S.~Mrenna, and P.~Z. Skands,
\newblock Comput. Phys. Commun. {\bf 178}, 852 (2008), 0710.3820.

\bibitem{Sjostrand:2014zea}
T.~Sj{\"o}strand {\em et~al.},
\newblock Comput. Phys. Commun. {\bf 191}, 159 (2015), 1410.3012.

\bibitem{Sirunyan:2018wem}
CMS, A.~M. Sirunyan {\em et~al.},
\newblock Phys. Rev. {\bf D97}, 112003 (2018), 1803.08856.

\bibitem{Carrazza:2019sec}
S.~Carrazza, C.~Degrande, S.~Iranipour, J.~Rojo, and M.~Ubiali,
\newblock Phys. Rev. Lett. {\bf 123}, 132001 (2019), 1905.05215.

\end{thebibliography}


\end{document}